\documentclass{webofc}
\usepackage[varg]{txfonts}   % Web of Conferences font
\usepackage{rotating}
\usepackage{tikz}
\graphicspath{{../../../Auxilliary/}}

\let\jnlstyle=\bf%rmfamily
\def \refjnl#1{{\jnlstyle#1}}

\newcommand\apj{\refjnl{ApJ}, }%
\newcommand\apjl{\refjnl{ApJ}, }%
\newcommand\apjs{\refjnl{ApJS}, }%
\newcommand\ao{\refjnl{Appl.~Opt.}}%
\newcommand\nat{\refjnl{Nature}, }%
\newcommand\pasp{\refjnl{PASP}, }%
\newcommand\aap{\refjnl{A\&A}, }%
\newcommand\mnras{\refjnl{MNRAS}, }%
\newcommand\zap{\refjnl{ZAp}, }%
\begin{document}
\title{Diffraction-dominated observational astronomy}

\author{\firstname{Frantz}
  \lastname{Martinache}\inst{1}\fnsep\thanks{\email{frantz.martinache@oca.eu}}
}

\institute{Laboratoire Lagrange, Université Côte d'Azur, Observatoire de la
  Côte d'Azur, CNRS.
}

\abstract{This paper is based on the opening lecture given at the 2017 edition
  of the Evry Schatzman school on high-angular resolution imaging of stars and
  their direct environment. Two relevant observing techniques: long baseline
  interferometry and adaptive optics fed high-contrast imaging produce data
  whose overall aspect is dominated by the phenomenon of diffraction. The
  proper interpretation of such data requires an understanding of the coherence
  properties of astrophysical sources, that is, the ability of light to produce
  interferences. This theory is used to describe high-contrast imaging in more
  details. The paper introduces the rationale for ideas such as apodization and
  coronagraphy and describes how they interact with adaptive optics. The
  incredible precision brought by the latest generation adaptive optics systems
  makes observations particularly sensitive to subtle instrumental biases that
  must be accounted for, up until now using post-processing techniques. The
  ability to directly measure the coherence of the light in the focal plane of
  high-contrast imaging instruments using focal-plane based wavefront control
  techniques will be the next step to further enhance our ability to directly
  detect extrasolar planets.}

\maketitle
\section{Introduction}
\label{intro}

This edition of the Evry Schatzman school is dedicated to the high angular
resolution imaging of the surface of stars and their direct environment. Two
families of observational techniques: adaptive-optics (AO) assisted
high-contrast imaging and long baseline interferometry, are contributing to
making this ambition a reality.

\begin{figure}[h]
  \hfill
  \includegraphics[height=6cm]{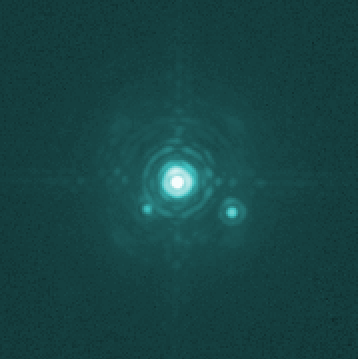}
  \hfill
  \includegraphics[height=6cm]{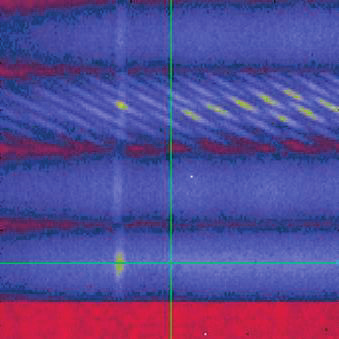}
  \hfill
  \caption{Two examples of diffraction-dominated data: On the left, a K-band
    AO-corrected image of the binary star $\alpha$-Ophiucus, observed from the
    Palomar Telescope. On the right, a set of spectrally dispersed
    two-telescope fringes produced by the instrument AMBER at the focus of
    VLTI. Both images are dominated by diffraction features such as fringes and
    rings and can also be affected by other artifacts like ghosts.}
  \label{f:diff}
\end{figure}

As different as they may seem at first look (see Figure \ref{f:diff}), the data
produced by these observational techniques share many characteristics. In both
cases, whether it be interference fringes or images boosted by a high-order AO
system, these data are dominated by diffraction features, that are the combined
signature of the telescope and instrumentation used to perform the
observations, and include the effect of ever changing atmospheric
conditions. The electromagnetic nature of the light collected by the
observatory, which can oftentimes be neglected when looking at wide-field
images, becomes manifest with these observing techniques since features such as
diffraction rings, fringes and speckles become prominent.

For each structure present in the data, one must be able to discriminate the
signature of a genuine structure like that of a faint planetary companion, a
clump in a circumstellar disk, or a structure of a stellar surface, from a
diffraction feature. The ultimate discrimination criterion has to do with the
degree of coherence of the structure in question.

Figure \ref{f:diff} presents two examples of diffraction dominated frames, one
produced by a single telescope, the other by an interferometer. In both cases,
the question one needs to examinate is: {\it is the object I am looking at a
  point source or did my frame capture the presence of more complex
  structures?} To figure out how to answer this question, we need to take a
closer look at the process of image formation.

\section{Images in astronomy}
\label{sec:images}

Images are the starting point of a lot of astronomical investigations. Even to
the non-expert, because the image is a direct extension of one's intimate sense
of sight, it offers rapid insights into complex situations. The image is the
place where an observer will (1) identify new sources, (2) measure their
position and brightness relative to a set of references and (3) follow their
evolution as a function of time, wavelength and polarization. From these
fundamental measurements, populating a multi-dimensional map $I(\alpha, \delta,
\lambda, P, t)$ function of position, wavelength, polarization and time, an
astronomer will improve his/her understanding (i.e. build a model) of a given
object or event, that tells the story of an open star cluster, a group of
galaxies, or that of a young planetary system, forming in the vicinity of a
nearby star. The fair and efficient interpretation of images is essential to a
wide range of scientific applications.

Be it in an actual imaging instrument, a spectrograph or an interferometer, the
image is first and foremost, a peculiar optical locus, where the photons coming
from a wide number of sources, and more or less uniformly distributed over the
collecting surface (the pupil) of one or more telescopes, find themselves
optimally segregated by geometric optics. It is possible to describe the result
of this photon segregation process as the result $\mathbf{I}$ of a convolution
product between two parts: one that is representative of the true distribution
of intensities describing the source noted $\mathbf{O}$; and one that describes
the instrumental response, that includes the properties of the atmosphere, the
telescope and all the optics encountered by the light before reaching the
detector. This instrumental response, called point spread function is noted
$\mathbf{PSF}$, such that:

\begin{equation}
  \mathbf{I} = \mathbf{O} \otimes \mathbf{PSF},
  \label{eq:convol}
  \end{equation}

\noindent
where $\otimes$ represents the convolution operation.

Much effort is devoted by telescope and instrument designers to reduce the
impact of the instrumental contribution on the end product. For a great deal of
astrophysical observations, the improvement is such that one can directly
identify the object $\mathbf{O}$ to the image $\mathbf{I}$ without really
paying attention to the $\mathbf{PSF}$. The photon segregation process occuring
in the image plane is however fundamentally limited by the phenomenon of
diffraction. The scaling parameter that rules this limitation is the ratio
between the wavelength of observation $\lambda$ and the characteristic
dimension of the aperture used to perform the observation (the diameter of a
single telescope, or the length of the interferometric baseline). To quickly
estimate the angular resolution provided by a telescope, the following quick
formula often comes in handy:

\begin{equation}
  \theta \approx 200 \times \frac{\lambda}{D},
  \label{eq:quick_reso}
\end{equation}

\noindent
where $\theta$ is the angular resolution in milli-arcseconds (mas), $\lambda$
the wavelength in microns and $D$ the diameter of the aperture in meters. One
can verify that a one-meter telescope observing in the visible ($\lambda =
0.5\, \mu$m) offers a 100 mas (0.1'') angular resolution, and that an 8-meter
telescope observing in the near-infrared ($\lambda = 1.6\,\mu$m) gets down to
40 mas.

Yet even in seemingly ideal observing conditions, the segregation of photons
provided by the image is often not sufficient in solving some important
problems such as: (1) the identification of faint sources or structures in the
direct neighborhood of a bright object: in this context, the faint source one
tries to detect is competing for the observer's attention with the diffraction
features of its host or (2) the discrimination of sources of comparable
brightness so close to each other that they are said non-resolved. Dealing with
these two similar situations is the object of this presentation on
diffraction-dominated observational astronomy.

\section{Coherence properties of light}
\label{sec:coherence}

Electromagnetic radiation still contributes today to the great majority of the
information collected by astronomical observatories that forms the basis of
astrophysics: the properties of images produced by astronomical instrumentation
can be described using the results of an early XIX$^{th}$ century physics
theory laid out by James Clerck Maxwell. Electromagnetic waves consist of
synchronized oscillations of electric and magnetic fields that propagate
through a medium at an actual velocity smaller or equal to $c$ (the speed of
light through a vacuum). The electric and magnetic fields are orthogonal to one
another so that one can specify the wave by keeping track of the electric field
alone, which simplifies the description. Note that this presentation will not
discuss polarization effects, a refinement that can be added later and won't
change the results and properties derived. Electromagnetic (and therefore
electric) waves, are solutions to Helmoltz's equation (also called the wave
equation):

\begin{equation}
  \nabla^2 \mathbf{E} - \frac{1}{c^2} \ddot {\mathbf{E}} = 0,
\end{equation}

\noindent
where $c$ represents the propagation speed of these waves, i.e. the speed of
light. Natural solutions to this equation are oscillating functions with the
following form:

\begin{equation}
  \mathrm{E}_{\nu}(t,x) = \mathrm{E}_0 e^{i(kx-\omega t)} = \mathrm{E}_0
  e^{i2\pi(x/\lambda-\nu t)},
  \label{eq:E_lin}
\end{equation}

\noindent
characterized by a frequency $\nu$ corresponding to the number of oscillations
of the field per seconds (or Hertz) and the wavelength $\lambda$ that
corresponds to the distance covered by the propagating electric field over the
time of one oscillation. In a vacuum, these two quantities are related via the
following inverse relation:

\begin{equation}
  \lambda = c/\nu.
\end{equation}

The complex exponential form of the oscillating solution of Eq. \ref{eq:E_lin}
allows to separate the time and space dependencies of the electric field. The
spatial component is awarded a special name: the complex amplitude, noted
$A(x)$ such that:

\begin{equation}
  \mathrm{E}_{\nu}(t,x) = \mathrm{A}(x) \: e^{-i2\pi\nu t}.
\end{equation}

While the complex amplitude is written as the function of a single variable
$x$, one has to keep in mind that this complex amplitude is
tri-dimensional. Thus if for instance, the origin of the electromagnetic field
is a single point source, the electric field is a spherical function of a
radius coordinate $r$:

\begin{equation}
  \mathrm{E}_{\nu}(t,r) = (1/r) \: \mathrm{E_0} e^{i(kr-\omega t)}.
\end{equation}

\begin{figure}
  \hfill
  \includegraphics[width=6cm]{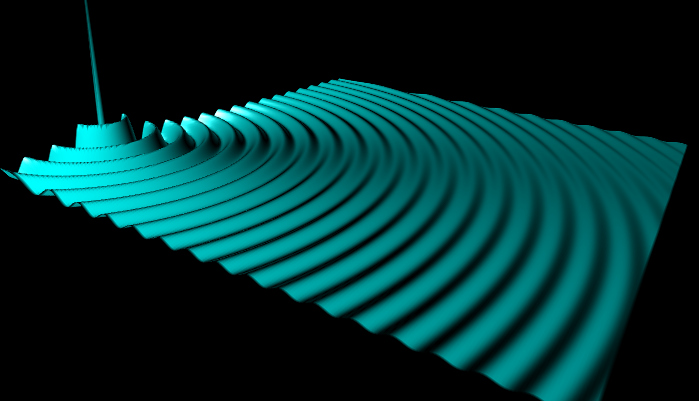}
  \hfill
  \includegraphics[width=6cm]{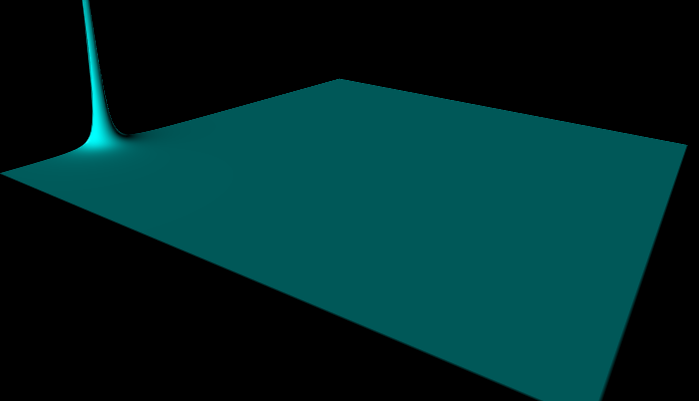}
  \hfill
  \caption{Propagation of a spherical wave. Left: imaginary instant snapshot of
    the complex amplitude of an electric field emitted by a point
    source. Directly observing this phenomenon would require a signal commuting
    time a million time shorter than state of the art fast switching
    semi-conductors can offer. Right: the static, time-averaged intensity
    associated to this same electric field, that can indeed be measured in the
    vicinity of a point source.}
  \label{f:spheric}
\end{figure}

The applications covered in this text relate to what is referred-to as the
optical: a regime of wavelength that covers the visible, going from $\lambda
\sim$ 0.4 $\mu$m to $\lambda \sim$ 0.8 $\mu$m where our human eye is mostly
sensitive and the infrared (IR) for wavelengths going up to $\lambda \sim$ 50
$\mu$m.  Beyond the infrared, it is customary to use the frequency $\nu$ to
describe the electromagnetic radiation. For wavelengths shorter than $\sim$ 100
nm, it is customary to use the energy associated to the radiation. Taking
$\lambda = 1 \mu$m as a wavelength representative of the optical and converting
it to a frequency:

\begin{equation}
    \nu = \frac{c}{\lambda} = 
  \frac{3\times10^8}{10^{-6}} = 3 \times 10^{14} \:\mathrm{Hz}.
\end{equation}

This really large number explains the specificity of the optical regime. The
typical read/write access time of today's fast switching semi-conductors is of
the order of $\sim$ 1 ns. Which means that over the time it takes to switch at
least once to take a snapshot, the electric field associated to optical light
has oscillated more than $10^5$ times. Unlike what is possible in the radio,
available readout electronics are not fast enough to record the value of the
electric field at any instant (see Figure \ref{f:spheric}). Instead, one
measures the time averaged energy carried by the field and intercepted by a
receiver, a quantity called the intensity:

\begin{eqnarray}
I \propto \langle |\mathrm{E}|^2 \rangle &=& \int_{t_0}^{t_0+\tau}
\mathrm{E}(t)^2\:\mathrm{d}t \\
&=& |\mathrm{A}|^2 \:\: (\mathrm{with}\: \tau >> 1/\nu).
\label{eq:intens}
\end{eqnarray}

\noindent
that is proportional to the square modulus of the complex amplitude. In the
absence of perturbations, the intensity recorded is a quantity that is only a
function (see Figure \ref{f:spheric}) of the relative spacing between the
source and the observer.

\begin{figure}
  \hfill
  \includegraphics[width=6cm]{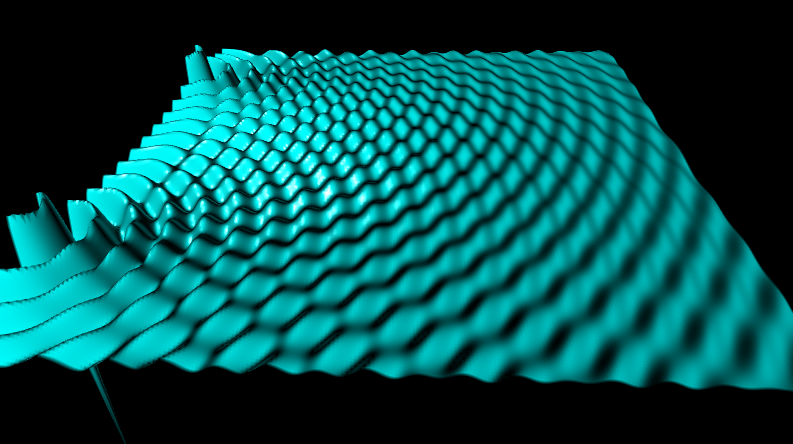}
  \hfill
  \includegraphics[width=6cm]{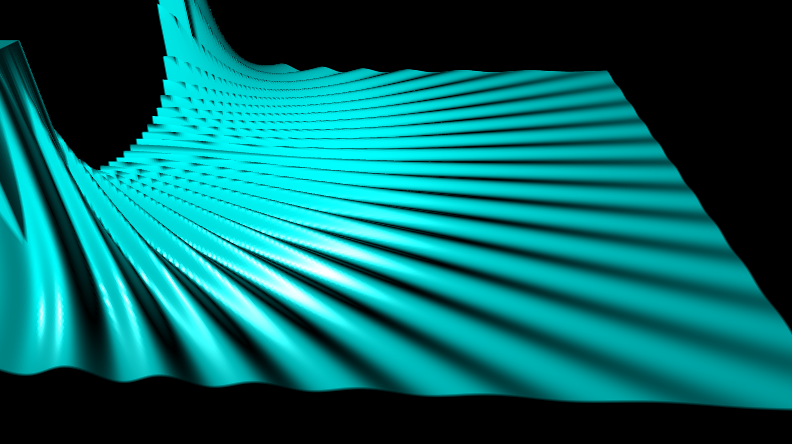}
  \hfill
  \caption{Visualisation of the interference phenomenon. Two point sources,
    located on the left hand side of the image at the same frequency, both emit
    a field propagating. Left: The two individual fields add up coherently and
    produce a rich wave pattern whose periodic properties depend on the
    frequency of the emission of the sources, and the distance that separates
    the two sources. Right: the static, time averaged intensity associated to
    this electric field. Unlike the single source scenario, in the far field
    (toward the right end of the image), intensity oscillations can be measured
    along the transverse direction.}
  \label{f:two_src}
\end{figure}

While apparently invisible when considering a single point source, the
oscillating nature of the electric field becomes manifest when when a second
source is present. The superposition principle states that the solution to this
new situation is the sum of the two individual fields. Figure \ref{f:two_src}
shows what one such field looks like. We still don't have a receiver fast
enough to be able to record the oscillations of the resulting field. The
intensity associated to the field (visible in the right panel of Figure
\ref{f:two_src}) however now also features some structure: the intensity
oscillates and depending on where the receiver is placed, one can either record
a maximum or a minimum of intensity. The distance between two consecutive
maxima of intensity will be a function of the ratio between the wavelength
$\lambda$ and the distance separating the two sources. Optical interferometry
is primarily concerned with the characterization of these structures, refered
to as interference fringes.

This mathematical description of the electromagnetic nature of light would
suggest that interference phenomena such as the one that was just described
should be commonplace. There are plenty of situations of every day life where
the light of two or more sources overlaps on a surface and yet, fringes are a
rare occurence. This is because our description has idealized the sources: the
purely sinusoidal wave (Eq. \ref{eq:E_lin}) is only suited to the description
of a laser beam.

The light emitted by thermal light sources like light bulbs or the hot gas of a
star originates from a large number of semi-random spontaneous and therefore
uncorrelated events like electronic transitions. A more accurate representation
of such an emission process uses a series of wave-packets, such as the ones
represented in Figure \ref{f:wave_packets}, which are a series of damped
oscillating fields $E_k$ modulated by an envelope function $\mathrm{env}$ and
characterized by a random emission time $t_k$ and a random phase at origin
$\Phi_k$:

\begin{figure}
  \includegraphics[width=\textwidth]{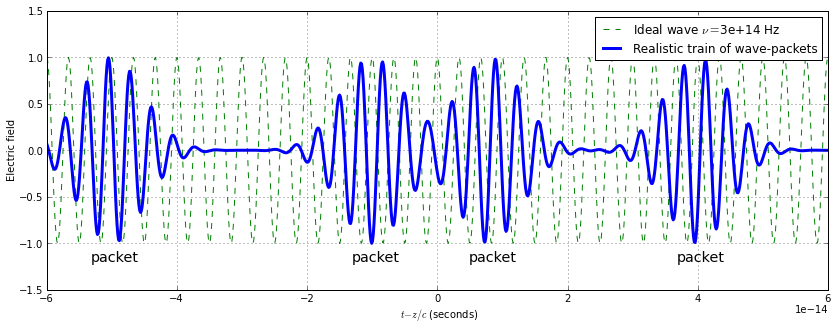}
  \caption{Improved, wave-packet based description of the electric field
    emitted by a thermal source (solid blue line) compared to the idealized
    sinusoidal field (dashed green line) used earlier. Random emission times
    and phase at the origin for each packet result in fluctuations in amplitude
    and phase of the electric field that will affect its capability to produce
    interferences.}
  \label{f:wave_packets}
\end{figure}

\begin{equation}
  \mathrm{E_k}(r,t) = \mathrm{env}(t-{t_k}) \times e^{i2\pi(r/\lambda-\nu
    (t-{t_k}) + {\Phi_k})}.
\end{equation}

The resulting electric field is no longer purely sinusoidal and fluctuates both
in its amplitude and phase: Figure \ref{f:wave_packets} compares this improved
wave-packet model to the earlier ideal wave and shows that these fields are no
longer synchronized, with the new electric field sometimes ahead of, and
sometimes behind the reference. This desynchronization will affect the capacity
of the light to produce interferences, a property characterized by a scalar
(complex) quantity called the degree of coherence.

The degree of coherence is the result of time-averaged cross-correlation
function. It can be used to compare and quantify how look-alike two distinct
electric fields are, in which case it will be referred-to as the mutual
coherence, or to compare one field with itself delayed in time, which will be
referred-to as the self-coherence. This self-coherence is a normalized complex
quantity:

\begin{equation}
  c(\tau) = \frac{<E^*(t) \times E(t+\tau)>}{<|E(t)|^2>},
\end{equation}

\noindent
whose modulus $|c(\tau)| \le 1$. In the case of the ideal wave model, the
degree of self-coherence is always equal to one: regardless of the time delay,
the electric field will always perfectly interfere with itself delayed in time.

\begin{figure}
  \includegraphics[width=\textwidth]{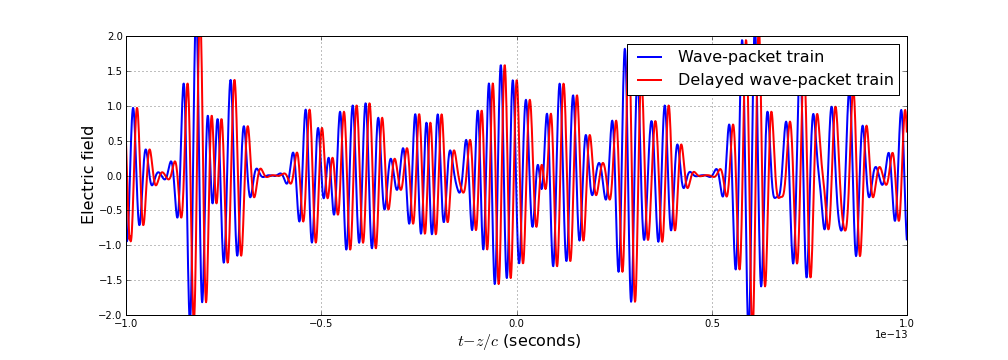}

  \includegraphics[width=\textwidth]{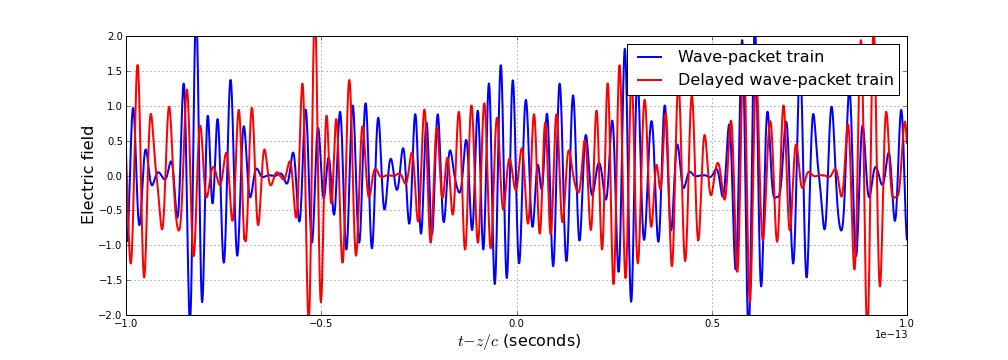}
  \caption{Illustration of the self-coherent properties of an electric field
    consisting of a collection of wave-packets. Top panel: when the time delay
    between the two is smaller than the coherence time $\tau_0$, the two fields
    do look alike and the result of their time-averaged cross-correlation
    exhibits a modulus $c \sim 1$. Bottom panel: beyond the coherence time, the
    two fields quickly decorrelate and the modulus of their self-coherence
    reaches $c \sim 0$.}
  \label{f:self-delay}
\end{figure}

In the wave-packed model, the field is only coherent with itself over when the
delay is small. Figure \ref{f:self-delay} presents two scenarios: a small delay
for which the original signal and its copy obviously correlate (ie. look
alike); and a large delay (larger than the size of one fringe packet) for which
the two fields clearly do not correlate anymore.

Nevertheless, even ni the second scenario, over a small range of time delays
bound by $\tau_0$ (the coherence time), one can measure reasonably good
correlation between the two signals. If one samples the same field twice, for
instance by placing holes in a screen equally distant from a point source, and
combines the two fields downstream such that their respective packets reach the
same place within the coherence time, interferences can be observed.

When one considers two fields $E_1$ and $E_2$ emanating from different sources,
one uses the degree of mutual coherence:

\begin{equation}
  c_{12} (\tau) = \frac{\langle E_1(t+\tau) E_2(t)^* \rangle} {\sqrt{\langle
      |E_1(t)|^2 \rangle \langle |E_2(t)|^2 \rangle}} = \frac{1}{\sqrt{I_1
      I_2}}\int\limits_{\Delta t} E_1(t+\tau) E_2^*(t) \:\mathrm{d}t
\end{equation}

\noindent
to quantify their capacity to interfere with one another. At this point, the
reader may have already guessed that two electric fields originating from two
distinct series of semi-random events have no chance of being correlated: the
expected mutual degree of coherence is equal to zero.

\begin{figure}
  \begin{tikzpicture}
    %\draw[help lines] (0,0) grid +(6,4);
    \definecolor{dylow}{RGB}{255,150,0};
    \definecolor{dgreen}{RGB}{0,100,0};
    \draw[black, fill=dylow!60, opacity=0.8]
    (5,2) ellipse [x radius=0.2cm, y radius=0.5cm];
    \node [draw=none] at (5,3) {R};
    
    \draw [red, fill=red] (0,2) circle (0.2cm);
    
    \draw [dgreen!60, fill=dgreen!60] (4,0) circle (0.2cm);
    \draw [dgreen!60, fill=dgreen!60] (4,4) circle (0.2cm);
    
    \node at (2,3) {\includegraphics[angle=25, scale=0.16]
      {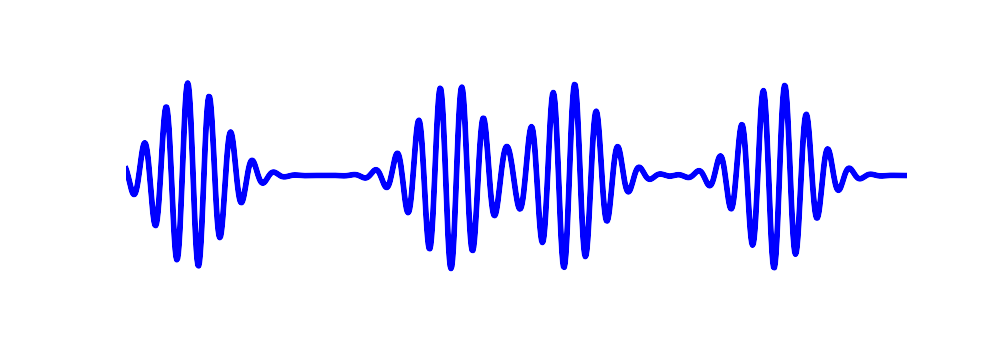}};
    \node at (2,1) {\includegraphics[angle=-25, scale=0.16]
      {wave_packets_1_naked.png}};
    
    \draw [<->,>=stealth] (0, 2.3) -- (4, 4.3) 
    node[above, midway] {$r_1$};
    
    \draw [<->,>=stealth] (0, 1.7) -- (4,-0.3)
    node[below, midway] {$r_2$};
    
    \draw [->,>=stealth] (4, 0) -- (5, 2);
    \draw [->,>=stealth] (4, 4) -- (5, 2);
    
    \node [draw=none] at (0, 2.5) {$S_1$};
    
    \node [draw=none] at (4.5, 4) {$P_1$};
    \node [draw=none] at (4.5, 0) {$P_2$};
    
    \node [draw=none,font=\LARGE] at (2.5, 2.5) {$E_1(t)$};
    \node [draw=none,font=\LARGE] at (2.5, 1.5) {$E_2(t)$};
    
    \draw [ultra thick] (-0.5,-0.5) -- (5.5,-0.5) -- (5.5, 4.5) -- (-0.5, 4.5)
    -- cycle;
    
    \node [draw=none] at (2.5,  5) {\small \bf self-coherence};
    \node [draw=none] at (2.5, -1)   {\small $S_1$: the source};
    \node [draw=none] at (2.5, -1.4) {\small $E_1$ and $E_2$: the electric fields};
    \node [draw=none] at (2.5, -1.8) {\small $P_1$, $P_2$: the observing stations};
    \node [draw=none] at (2.5, -2.2) {\small $R$: mono-pixel quadratic detector};
  \end{tikzpicture}
  \hfill
  \begin{tikzpicture}
    \definecolor{dylow}{RGB}{255,150,0};
    %\draw[help lines] (-1,-2) grid +(4,6);
    \draw[black, fill=dylow!60, opacity=0.8]
    (4,2) ellipse [x radius=0.2cm, y radius=0.5cm];
    \node [draw=none] at (4,3) {R};
    
    \draw [dylow, fill=dylow] (0,0) circle (0.2cm);
    \draw [red, fill=red] (0,4) circle (0.2cm);
    
    \node at (2,3) {\includegraphics[angle=-25, scale=0.16]
      {wave_packets_1_naked.png}};
    \node at (2,1) {\includegraphics[angle=+205, scale=0.16]
      {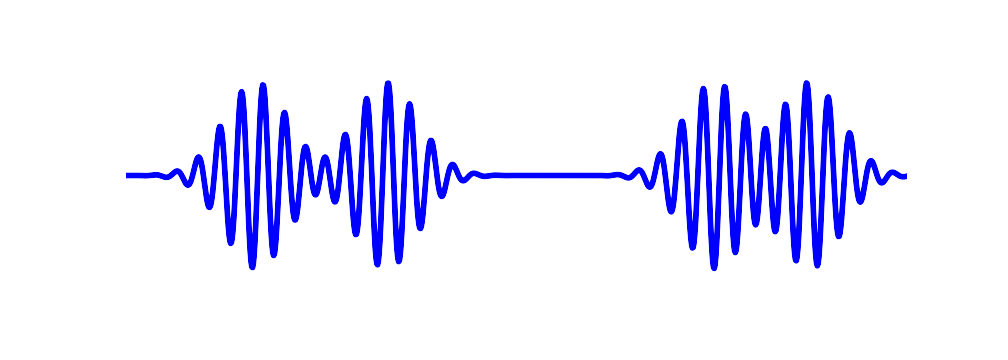}};
    \node [draw=none] at (0, 3.5) {$S_1$};
    \node [draw=none] at (0, 0.5) {$S_2$};
    
    \node [draw=none,font=\LARGE] at (1.5, 2.5) {$E_1(t)$};
    \node [draw=none,font=\LARGE] at (1.5, 1.2) {$E_2(t)$};
    
    \draw [ultra thick] (-0.5,-0.5) -- (4.5,-0.5) 
    -- (4.5, 4.5) -- (-0.5, 4.5) -- cycle;
    
    \node [draw=none] at (2,  5) {\small \bf spatial incoherence};
    \node [draw=none] at (2, -1) {\small $S_1$ and $S_2$: the two sources};
    \node [draw=none] at (2, -1.4) {\small $E_1$ and $E_2$: the electric fields};
    %\node [draw=none] at (2, -1.8) {\small $R$: mono-pixel quadratic detector};
    \node [draw=none] at (2, -2.2) {\small $R$: mono-pixel quadratic detector };

  \end{tikzpicture}
  \caption{Self- and spatial-coherence properties of the light emitted by
    astrophysical sources. Left: The light emitted by one {\bf point} source,
    collected by two or more apertures, and recombined in a manner that
    guarantees that all path lengths are equal (from the source to the
    detector), will lead to perfect interferences: {\bf point sources are
      self-coherent}. Right: The light from distinct astronomical sources, be
    it two distinct objects or two parts of the surface of one object, does not
    interfere: {\bf astronomical sources are spatially incoherent}.}
  \label{f:recap_coh}
\end{figure}

These two elementary observations on the self- and the mutual-coherence of the
electromagnetic fields emanating from thermal sources explain all the
properties of the formation of image and interference fringes in astronomical
instrumentation. Figure \ref{f:recap_coh} offers a graphical summary of these
two properties, and leads to the formulation of two simple but very important
facts about the light of natural light sources:

\begin{itemize}
\item Fact \#1: the light emitted by one {\bf point} source, collected by two
  or more apertures (or parts of one aperture), and recombined in a manner that
  all path lengths are equal, will lead to perfect interferences. {\bf
    Unresolved point sources are self-coherent}.
\item Fact \#2: the light from distinct astronomical sources, either distinct
  objects or two parts of the surface of one object, does not interfere. {\bf
    Astronomical sources are spatially incoherent}.
\end{itemize}

Most of the observing scenarios we are interested in in this lecture focus on a
bright, unresolved object that, in most cases, can be treated like a bright
point source, surrounded by fainter structures, such as planetary companions, a
dust shell or elements of a circumstellar disk.

In such a situation, the effective resulting coherence will be dominated by the
coherence properties of the bright source, but will be reduced due to the
presence of faint sources whose light does not interfere with that of the
bright source. The light of a single point source is perfectly coherent: in the
case of an interferometer, the estimator of coherence, called the fringe
visibility (or the visibility squared), is also equal to one; in the case of a
single telescope observation, the image consists of a single, crisp PSF. The
presence of additional structures around the bright point source will reduce the
apparent visibility of the fringes (in the case of the interferometer) and/or
make the single telescope image look fuzzier than on the point source alone:
the effective coherence of one such extended source takes intermediate values
between 0 and 1.

Being able to measure the coherence of a source from an interferogram or an
image assumes that one perfectly knows what the PSF or the fringe pattern
acquired on a point source actually looks like. It turns out that several
instrumental and environmental effects like the spectral bandpass, atmospheric
dispersion, residual aberrations or drifts can result in an apparent loss of
coherence. The task is somewhat easier when interpreting a two-aperture
interferogram, since the interferometer is really designed to produce
unambiguous measurements of coherence, than from an image that contains a
complex mix of overlapping spatial frequencies. The deconvolution of images,
that is the inversion of Eq. \ref{eq:convol}, is in practice difficult when the
PSF is not perfectly known as the problem is degenerate. We will get back to
this very question toward the end of this presentation and see how we can
addressit and make our coherence estimates unambiguous.

\section{Diffraction-dominated imaging}
\label{sec:diff}

Since another lecture specifically deals with interferometry, the discussion
will from now focus on the properties of images produced by single
telescopes. Hopefully, the reader will realize that it doesn't take long to
adapt the following discussion to the case of a multi-aperture interferometer.

\def\x0{1.0}
\def\y0{1.0}
\definecolor{dgreen}{RGB}{0,100,0}

\begin{figure}
  \centering
\begin{tikzpicture}
  %\draw[help lines] (0,0) grid +(10,4);

  \draw [thick] (-0.5,-0.5) -- (10.5,-0.5) 
  -- (10.5, 4.7) -- (-0.5, 4.7) -- cycle;

  \draw [red, fill=red!40]
  (1.5, 0.5) .. controls (0.5,0.5) and (0.5,1.5) .. (1,2.0) -- 
  (1.5, 3.0) .. controls (2.5, 3) .. (2.5,2) -- 
  (2,1.5) .. controls (2,0.5) .. (1.5,0.5);

  \draw [thick, <->, >=stealth] (1.5, -0.2) -- (8.5, -0.2) 
  node [midway, above] {$Z$};

  \draw [dashed, thick ] (0.0, 2.0) -- (10.0, 2.0);
  \draw [ultra thick, ->,>=stealth] (1.5, 0.0) -- (1.5, 4.0) node
        [above] {Y};
  \draw [ultra thick, ->,>=stealth] (3.0, 3.0) -- (0.0, 1.0) node
        [above] {X};

  \draw [ultra thick, ->,>=stealth] (8.5, 0.0) -- (8.5, 4.0) node
        [above] {y};
  \draw [ultra thick, ->,>=stealth] (7.0, 3.0) -- (10., 1.0) node
        [above] {x};

  \filldraw [dgreen!60] (7.5,1.5) ellipse [x radius=0.1cm, y radius=0.15cm]; 

  \filldraw [red] (\x0,\y0) ellipse [x radius=0.1cm, y radius=0.15cm];

  \draw [thick, <->,>=stealth]  
    (\x0,\y0) -- (7.5, 1.5) node [above, midway] {$r$};

  \node [draw=none] at (7.8, 1.8) {$M(x,y)$};

  \node [red, draw=none] at (2.0, 3.2) {$\Sigma$};
  \node [red, draw=none] at (\x0-0.4, \y0-0.3) {d$\sigma$};
  \node [draw=none] at (\x0-0.4, \y0-0.8) {P(X,Y)};
\end{tikzpicture}
\caption{Schematic representation of the phenomenon of diffraction. A diaphragm
  described by the support $\Sigma$, on the left-hand side is uniformly lit by
  a point source at infinity. The symbols and notations present in this figure
  are used to determine the distribution of complex amplitude $A(x,y)$ in the
  right hand side plane located at a distance $Z$ from the diaphragm.}
\label{f:diffraction}
\end{figure}
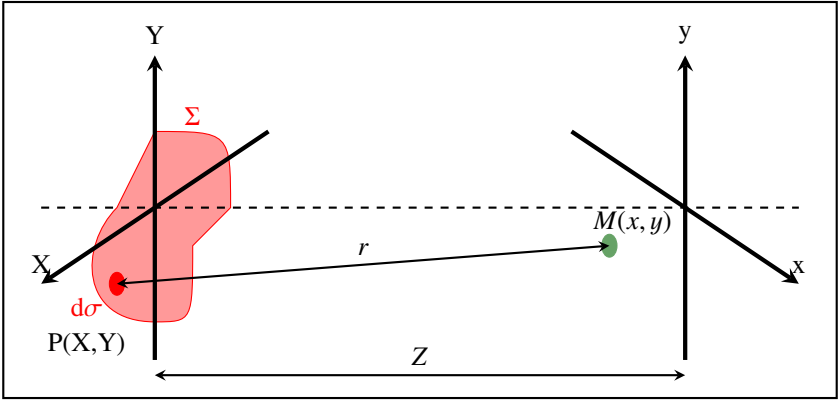

Figure \ref{f:diffraction} introduces the symbols and the scenario used to
describe the phenomenon of diffraction: on the left, a diaphragm or arbitrary
shape described by the surface $\Sigma$, uniformly lit by a point source
located so far toward the left that (under perfect observing conditions) the
complex amplitude of the associated electric field is constant across the
aperture.  $\Sigma$ describes the aperture of the telescope used to do
imaging. If one were to consider looking into interferometry from here, one
would just have to split $\Sigma$ into a collection of sub-apertures.

The important relation to establish is one that relates the electric field (at
least its complex amplitude) across the aperture $\Sigma$ to its counterpart
projected on a screen located at a distance Z from the diaphragm. One
elementary surface element d$\sigma$ is singled out on this picture. This
elementary surface element is the origin of a new spherical wave (a principle
described by Augustin Fresnel in 1818). For a point M of coordinates $(x,y)$
located at a distance r from the origin of the wave, the contribution for the
wavelength $\lambda$ to the local complex amplitude from d$\sigma$ is given by:

\begin{equation}
  \mathrm{d}A(x,y) = \frac{1}{r} \times K \times A(X,Y) \times e^{j2\pi r /\lambda} \, \mathrm{d}\sigma,
\end{equation}

\noindent
where K is a constant. Since we've established that the light associated to a
single point source is coherent, we can write that the total electric field in
right-hand side plane of Figure \ref{f:diffraction} is the result of a sum of
emissions from all elementary point sources:

\begin{equation}
  A(x,y) = K \iint_\Sigma \frac{1}{r} \times A(X,Y) \times e^{j2\pi r /\lambda} \, \mathrm{d}\sigma.
\end{equation}

If the distance Z between the two diaphragms and the backend screen is
sufficiently large in comparison to the dimension of the diaphragm, the
distance $r$ can be approximated:

\begin{eqnarray}
r &=& \sqrt{Z^2 + (X-x)^2 + (Y-y)^2} \\
&\approx& Z \biggl(1 + 0.5\biggl(\frac{X-x}{Z}\biggr)^2 + 
                       0.5\biggl(\frac{Y-y}{Z}\biggr)^2 \biggr).
\end{eqnarray}

So that the expression for the complex amplitude in the plane on the right hand
side of Figure \ref{f:diffraction} can be rewritten as the result of:

\begin{equation}
  A(x,y) = \frac{K}{Z}e^{i2\pi Z/\lambda} \iint_\Sigma A(X,Y) 
  \exp{\biggl(\frac{i\pi}{\lambda Z}((X-x)^2+(Y-y)^2)\biggr)} 
  \,\mathrm{d}\sigma.
  \label{eq:fresnel}
\end{equation}

This form of integral is called the Fresnel Transform. It is a non-linear
transform whose computation can therefore be a bit cumbersome. It is however
very general and can be used to compute the diffraction by a diaphragm for a
wide range of situations. The Fresnel Transform of Eq. \ref{eq:fresnel}
can however be further simplified when the distance $Z$ between the diaphragm
and the screen becomes very large, compared to the dimensions of the aperture:

\begin{equation}
  \exp{\biggl(\frac{i\pi}{\lambda Z}(X-x)^2\biggr)} \approx \exp{\biggl(\frac{i\pi}{\lambda Z}x^2\biggr)} \times \exp{\biggl(\frac{-i2\pi}{\lambda Z}x X \biggr)},
\end{equation} 

\noindent
when $\frac{X^2}{\lambda Z} << 1$. This situation is referred-to as the far
field or the Fraunhofer diffraction. While it seems like an approximation, it
is perfectly suited to the description of what is happening when a powered
optics (see Figure \ref{f:focus}) is used to conjugate an object, located at
infinity, to its image, placed at a finite distance. In the focal plane of a
telescope, the far field approximation becomes perfectly valid. The Fresnel
Transform of Eq. \ref{eq:fresnel} can be rewritten as:

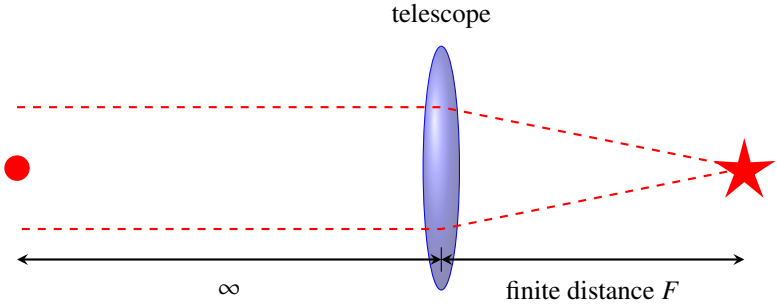
\begin{figure}
  \centering
  \begin{tikzpicture}[scale=0.8]
    \draw[blue]
    (7,0) ellipse [x radius=0.3cm, y radius=2cm];
    % --
    \shade[ball color=blue!80!white,opacity=0.50] 
    (7, 0) ellipse [x radius=0.3cm, y radius=2cm];
    % --
    \draw[dashed, red, thick] (0, 1) -- (7, 1) -- (12, 0) -- (7, -1) -- (0, -1);
    % --
    \draw [<->,>=stealth, thick] (0, -1.5) -- (7, -1.5);
    \draw [<->,>=stealth, thick] (7, -1.5) -- (12, -1.5);
    \draw (7, -1.7) -- (7, -1.3);
    % --
    %\node [draw=none, scale=4] at (0,0) {\color{red}$\star$};
    \draw [red, fill=red] (0,0) circle (2mm);
    
    \node [draw=none, scale=4] at (12,0) {\color{red}$\star$};
    \draw (7,  2.5) node {telescope};
    \draw (3.5,   -2) node {$\infty$};
    \draw (9.5, -2) node {finite distance $F$};
\end{tikzpicture}
\caption{The impact of geometric optics on diffraction: a powered optical
  element conjugates an object located at infinity to an image at a finite
  distance in the focal plane. To compute the effect of the diffraction by the
  telescope in the focal plane, one can safely use the Fraunhofer diffraction
  relying on the computation of Fourier Transforms. To compute the effect of
  diffraction at any plane located in between the image and the diffractive
  aperture, one must use the more general Fresnel diffraction.}
\label{f:focus}
\end{figure}

\begin{equation}
A(x,y) = K' \iint_\Sigma A(X,Y) \exp{\biggl(\frac{-i2\pi}{\lambda F} (xX +
  yY)\biggr)} \, \mathrm{d}\sigma.
\label{eq:fraunhofer}
\end{equation} 

\noindent
where the distance $Z$ has been replaced by the focal length $F$ of the imaging
optics. It is convenient to express the coordinates in the image in terms of
angular distances relative to the pointing axis, replacing the ratio $x/F$ and
$y/F$ by angular coordinates $\alpha,\beta$. One can drop the $K'$ constant as
well to simplify the notations and just ensure in the computation that the
total number of photons collected during an integration, is preserved by the
transformation:

\begin{equation}
A(\alpha,\beta) = \iint_\Sigma A(X,Y) \exp{\biggl(\frac{-i2\pi}{\lambda}
  (\alpha X + \beta Y)\biggr)} \, \mathrm{d}X\mathrm{d}Y.
\label{eq:fourier}
\end{equation} 

\noindent
which you may recognize as the two dimensional Fourier Transform of the
distribution of the complex amplitude in the diffracting aperture. Unlike the
Fresnel Transform, the Fourier Transform (hereafter represented by the symbol
$\mathcal{F}$ is a linear operation that can be computed in an efficient
manner. This is the form we will mostly use for the rest of the cases described
in this lecture.

Equipped with this quantitative description of the diffraction and our previous
observations on the coherent properties of astronomical sources, we can outline
a recipe for the formation of an image:

\begin{enumerate}
\item an extended source can be described as a finite discrete collection of
  self-coherent point sources. The object function can be written as
  $O = \sum_k O_k$.
\item each point source uniformly illuminates the diffractive aperture. On
  axis, the complex amplitude ($A_p$) is constant. The complex amplitude of
  each off-axis source includes a phase slope that is proportional to how far
  off-axis that source is.
\item because each point source is perfectly self-coherent, in the focal plane,
  the complex amplitude $A_{f,k}$ associated to each point source is the result
  of the Fourier Transform of the complex amplitude of the field intercepted by
  the aperture: $A_{f,k} = \mathcal{F}(A_{p,k})$.
\item a detector only records the intensity associated to this point source:
  $I_k = |\mathcal{F}(A_{p,k})|^2$. The effect of the phase slope associated to
  the off-axis source of index $k$ translates the resulting intensity pattern.
\item due to the spatial incoherence property of astronomical sources, the
  intensity patterns of all point sources add up: $I = \sum_k I_k$.
\end{enumerate} 

\begin{figure}
  \tikzstyle{arrow} = [ultra thick,->,>=stealth]

  \centering
  \begin{tikzpicture}[scale=0.8]

    %\draw[help lines] (0,0) grid +(13,5);
    %\draw [thick] (0.0, 0.0) -- (13.0,0.0)  -- (13, 5.0) -- (0.0, 5.0) -- cycle;

    \node at (2.5, 2.5) {\includegraphics[width=4cm]{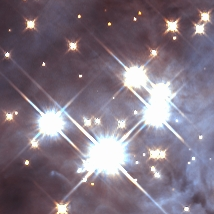}};

    \def\x0{2.5}
    \def\y0{1.4}
    
    \draw [red, ultra thick] (\x0-1, \y0-1) -- (\x0+1, \y0+1) -- (\x0, \y0) 
    -- (\x0-1, \y0+1) -- (\x0+1, \y0-1);
    \draw [red, fill=red] (\x0,\y0) circle (0.4cm);
    \draw [red, fill=red] (\x0 + 4.5,\y0) circle (0.2cm);

    \def\x0{1.2}
    \def\y0{2.1}
    \draw [red, ultra thick] (\x0-1, \y0-1) -- (\x0+1, \y0+1) -- (\x0, \y0) 
    -- (\x0-1, \y0+1) -- (\x0+1, \y0-1);
    \draw [red, fill=red] (\x0,\y0) circle (0.4cm);
    \draw [red, fill=red] (\x0 + 4.5,\y0) circle (0.2cm);

    \def\x0{3.6}
    \def\y0{2.4}
    \draw [red, ultra thick] (\x0-1, \y0-1) -- (\x0+1, \y0+1) -- (\x0, \y0) 
    -- (\x0-1, \y0+1) -- (\x0+1, \y0-1);
    \draw [red, fill=red] (\x0,\y0) circle (0.4cm);
    \draw [red, fill=red] (\x0 + 4.5,\y0) circle (0.2cm);

    \def\x0{3.8}
    \def\y0{2.8}
    \draw [red, ultra thick] (\x0-1, \y0-1) -- (\x0+1, \y0+1) -- (\x0, \y0) 
    -- (\x0-1, \y0+1) -- (\x0+1, \y0-1);
    \draw [red, fill=red] (\x0,\y0) circle (0.4cm);
    \draw [red, fill=red] (\x0 + 4.5,\y0) circle (0.2cm);

    \def\x0{3.2}
    \def\y0{3.2}
    \draw [red, ultra thick] (\x0-1, \y0-1) -- (\x0+1, \y0+1) -- (\x0, \y0) 
    -- (\x0-1, \y0+1) -- (\x0+1, \y0-1);
    \draw [red, fill=red] (\x0,\y0) circle (0.4cm);
    \draw [red, fill=red] (\x0 + 4.5,\y0) circle (0.2cm);

    \node [black] at (2.5, 0.5) {\bf the image};
    \node [black] at (7.0, 0.5) {\bf the object};
    \node [black] at (11.5, 0.5) {\bf the PSF};

    \node [red] at (5.0, 2.5) {\LARGE =};
    \node [red] at (9.5, 2.5) {$\bigotimes$};

    \def\x0{11.5}
    \def\y0{2.5}
    \draw [red, ultra thick] (\x0-1, \y0-1) -- (\x0+1, \y0+1) -- (\x0, \y0) 
    -- (\x0-1, \y0+1) -- (\x0+1, \y0-1);
    \draw [red, fill=red] (\x0,\y0) circle (0.4cm);

  \end{tikzpicture}
  \caption{The image-object convolution relation illustrated: it is because of
    the spatial incoherence properties of astronomical sources present in the
    field (here extracted from a HST/NICMOS image of the Trapezium), that the
    image can be written as the result of a convolution product between the
    object function $O$ and the PSF of the telescope and its instrument.}
  \label{f:imgform}
\end{figure}
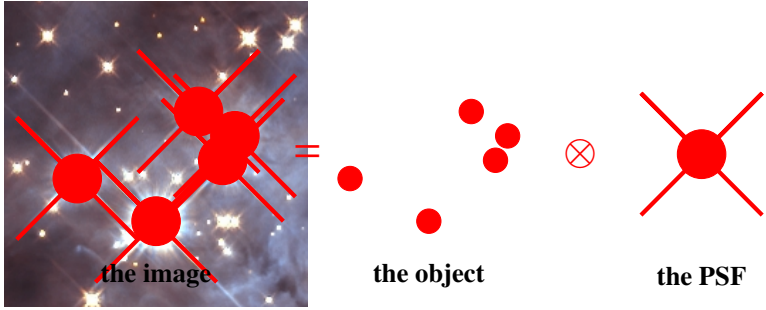

Since the light associated to each source is intercepted by the same aperture,
the shape of the intensity pattern associated to each source (i.e. the PSF) is
the same: the PSF is translation-invariant\footnote{When a single diffractive
  element is present only. In practice, the atmosphere, the relay optics inside
  the telescope and the instrument can render the PSF no-longer translation
  invariant. Over the small field of view we are dealing with here, these
  subtleties can be neglected.}. It is only modulated by the brightness of
individual sources that acts as a scaling factor. The image can therefore be
formally described as the weighted sum of PSFs. Figure \ref{f:imgform}
illustrates this property, which was given as early as Eq. \ref{eq:convol} in
this presentation but that we can now explain as the direct consequence of the
coherence properties of astronomical sources.

Given the importance of the PSF in the shaping of the final image (see Figure
\ref{f:imgform}), we need to see how the shape and size of the aperture, also
known as the pupil, will impact the PSF. The theory of diffraction outlined
earlier showed that the PSF can conveniently be computed as the result of the
square modulus of the Fourier Transform of the illumination of the pupil:

\begin{equation}
\mathrm{PSF} = \biggl|\mathcal{F}(\mathrm{pupil})\biggr|^2.
\end{equation}

Real telescopes unfortunately have fairly complex pupils, featuring at least a
central obstruction induced by the presence of a secondary mirror and spider
vanes that give support to this secondary mirror. The primary mirror itself can
also be made of several segments whose edges induce further diffraction. The
PSF of a circular unobstructed telescope (known as the Airy function), only
relevant for on-axis refractive telescopes or for an off-axis reflective one,
is a useful reference to compare a real telescope to. The circular aperture is
one of the few geometries for which the PSF has an analytical expression. Its
radial profile is described by:

\begin{equation}
\mathrm{Airy}(r) = 4 \times \biggl|\frac{J_1(\pi r)}{\pi r}\biggr|^2,
\end{equation} 

\noindent
where $r$ is the angular distance expressed in units of the ratio between the
wavelength and the diameter of the aperture ($\lambda/D$) and $J_1$ a Bessel
function. This Airy pattern, represented in Figure \ref{f:airy} (using a
logarithmic scale) features diffraction rings that extend very far away from
its core. The Airy function meets its first zero for $r = 1.22 \lambda/D$,
which is often used to estimate the order of magnitude for the angular
resolution of an observing setup. Regardless of the details of the aperture,
the ratio $\lambda/D$, where $D$ is the characteristic dimension of the
diffractive system\footnote{It can be the diameter of a single telescope or the
  distance separating two sub-apertures when dealing with interferometry.},
will always be the right order of magnitude to consider to characterize the
angular resolution of an optical setup. For an 8-meter diameter telescope
operating in the near infrared, the ratio $\lambda/D$ is of the order of
$ \sim 10^{-7}$ radians. Such a small value makes the radian a inconvenient
unit to manipulate. In practice, instrument plate scales for imagers at the
focus of space-borne or ground based AO-corrected telescopes are usually
expressed in milli-arc seconds per pixel. The conversion from radians to
arcseconds given by:

\begin{eqnarray*}
  \theta\: [''] &=& \frac{180 \times 3600}{\pi} \times \theta \: [\mathrm{rad}] \\
  &\simeq& 206264.8 \times \theta \: [\mathrm{rad}],
\end{eqnarray*}

\begin{figure}
  \centering
  \begin{tikzpicture}
    %\draw [fill=red]
    %(0,0) -- (4,0) -- (4,2) -- cycle;
    \draw [color=black, fill=red, fill opacity=0.6, line width=2pt]
    (0,0) -- (4,0) -- (4,2) -- cycle;
    \draw[black, line width=2pt] (0:1.0) arc (0:25:1.0);
    
    \draw [<->,>=stealth] (4.5,0.0) -- (4.5, 2.0);
    \draw(2.0, 0.4) node {angle $\theta$};
    \draw(2.0,-1.0) node {distance $d$ (pc)};
    \draw [<->,>=stealth] (0.0,-0.5) -- (4.0, -0.5);
    \draw(5.2, 1.0) node {1 AU};
  \end{tikzpicture}
  \caption{The parsec is the distance at which a projected distance of 1 AU
    (astronomical unit) corresponds to an angular distance of one arc
    second. One parsec is therefore roughly equal to $2 \times 10^5$ AU (see
    text).}
  \label{f:206265}
\end{figure}
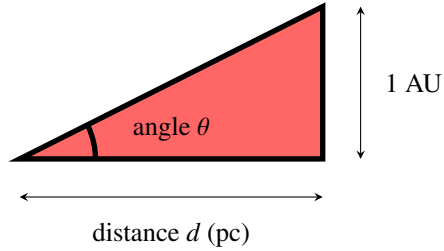 

\noindent
gave us the short-hand formula of Eq. \ref{eq:quick_reso} for the angular
resolution in mas. The 206,264.8 conversion factor (often rounded to
$2 \times 10^5$) is an order of magnitude that is good to keep in mind. It is
indeed the scaling factor between phenomena occuring inside a planetary system
(where distances are measured in astronomical units or AU) and phenomena
occuring over interstellar distances (for which distances are measured in
parsecs). Since the parsec was defined as the distance at which a projected
distance of 1 AU corresponds to an angle of one arc second (see
Fig. \ref{f:206265}):

\begin{eqnarray*}
  \tan{1 ''} \sim 1'' &=& 1\: \mathrm{AU} / 1\: \mathrm{pc} \\
  \theta\: [''] &=& 1 /d \: [\mathrm{pc}] \\
  1 \mathrm{pc} &=& 206264.8 \:\mathrm{AU}.
\end{eqnarray*}

%comment on the size and the shape of the diffractive aperture.

\section{High-contrast imaging}
\label{sec:hicont}

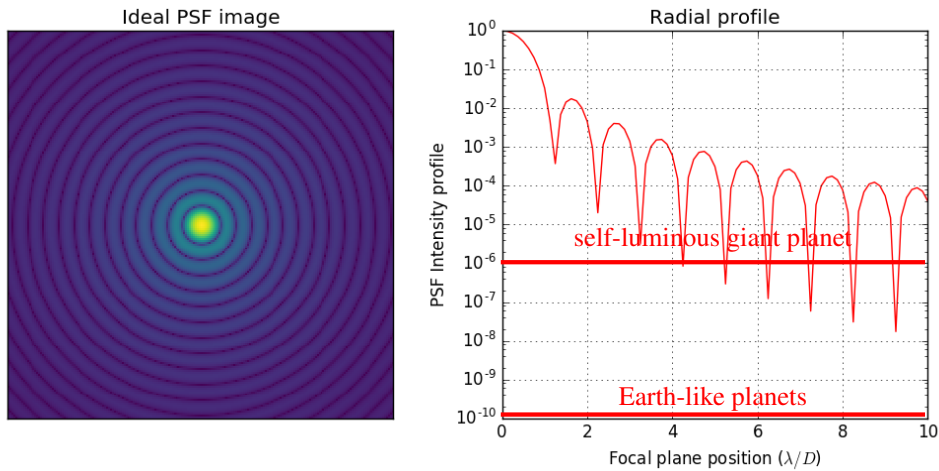
\begin{figure}
  \begin{tikzpicture}

    \node at (6.5, 3.5)
          {\includegraphics[width=13cm]{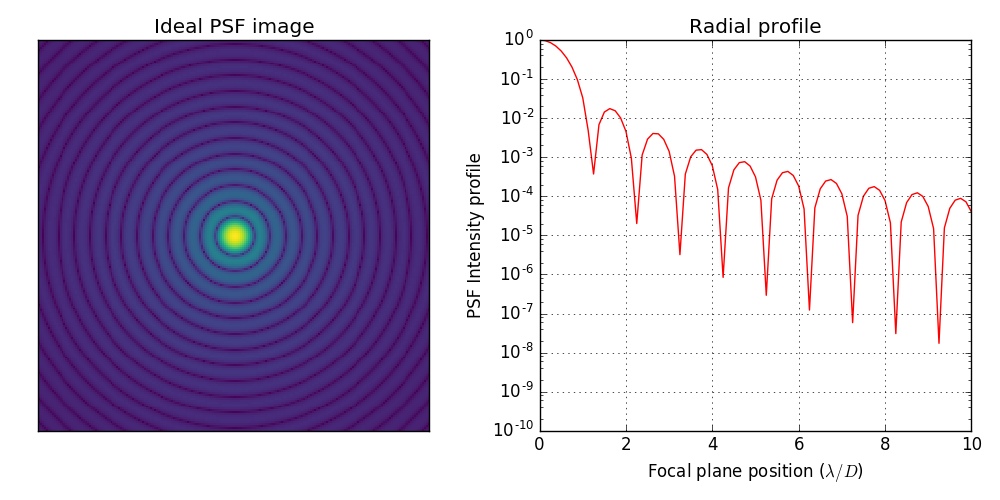}};
    %\draw[black] (0,0) grid +(13,7);
    
    \draw [red, ultra thick] (7.0, 3.2) -- (12.6, 3.2); 
    \node [red] at (9.8, 3.5) {self-luminous giant planet};
    
    \draw [red, ultra thick] (7.0, 1.2) -- (12.6, 1.2); 
    \node [red] at (9.8, 1.4) {Earth-like planets};

  \end{tikzpicture}
  \caption{Left: 2D representation of a the perfect point spread function (PSF)
    of a circular aperture (using a non-linear color scale). Right: radial
    profile of the same PSF, over a 10 $\lambda/D$ range of angular
    separation. Two horizontal lines mark the expected relative brightness
    (contrast) of two types of planets: self-luminous Jupiter-like at the
    $10^{-6}$ level and reflective Earth-like at the $10^{-10}$ level. Both
    lines lay several orders of magnitude below the PSF.}
  \label{f:airy}
\end{figure}

From the rather large sample of extrasolar planets known at the time of this
writing, only a dozen systems featuring planetary candidates have been imaged
by space-borne and ground-based telescopes. Why is this task so difficult?

For a nearby planetary system, i.e. located $\sim$20 parsecs away from our own
Solar system, planets on orbital distances between 1 and 10 AU will appear at
angular separations ranging from 50 mas to $0.5''$ which seems to be within the
angular resolution reach of modern telescopes, even when observing in the
near-infrared. The difficulty in the direct imaging of extrasolar planets lies
in the very large difference of luminosity between a faint planet and its
bright host star. The brightness ratio, also known as the contrast ratio, of a
mature Earth-like planet in a 1 AU orbit around an equally mature Sun-like star
would be characterized by an incredibly large $10^{-10}$ contrast ratio. A more
favorable scenario is that of a self-luminous giant planet like Jupiter in
orbit around a young star for which the contrast ratio could stay as high as
$10^{-6}$ for a few million years. The right panel of Figure \ref{f:airy}
illustrates the difficulty of the situation, by comparing these two scenarios
to the ideal PSF profile of a circular aperture. Even at the largest plotted
angular separation (10 $\lambda/D$), the signal one would like to detect is
still orders of magnitude fainter than the photon noise of the local
diffraction structures of the on-axis bright star. When the pupil of the
telescope features additional structures such as a central obstruction and
spider vanes, the situation is even less favorable.

Simply masking out the PSF in the focal plane does not contribute much: it can
help avoid saturation problems on the brightest parts of the PSF but the photon
noise of the light present in the diffraction rings will still be the dominant
source of noise. To facilitate the detection of faint structures present in the
neighborhood of a bright object, one needs to reduce overall on-axis
transmission so as to reduce the bright object's photon noise. The need for
high-contrast imaging gave birth to a wide number of techniques amongst which
two major families emerge: apodization and coronagraphy. Since the early 2000s,
this still active area of research has generated a lot of enthusiasm and become
extremely sophisticated. The goal of this presentation is not to give the
readers a detailed description of the state of the art, but rather to introduce
the important ideas that will help understand how the challenge can be
addressed. This will require the application of the diffraction theory that was
described earlier.

\subsection{Pupil apodization}
\label{sec:apod}

\begin{figure}
  \includegraphics[width=\textwidth]{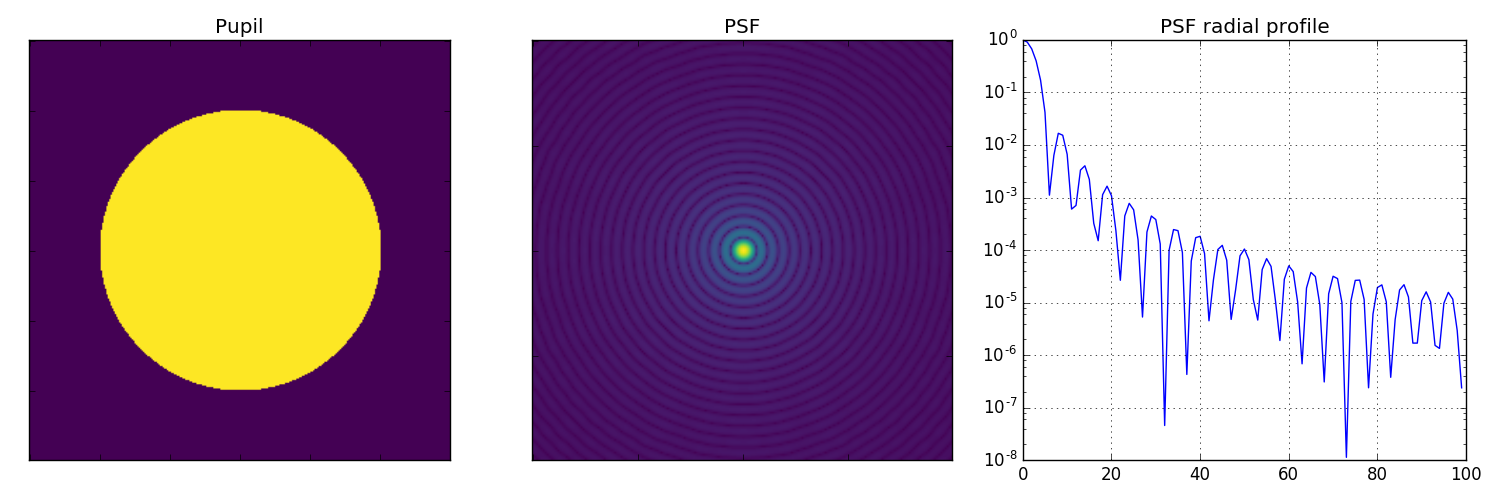}
  \includegraphics[width=\textwidth]{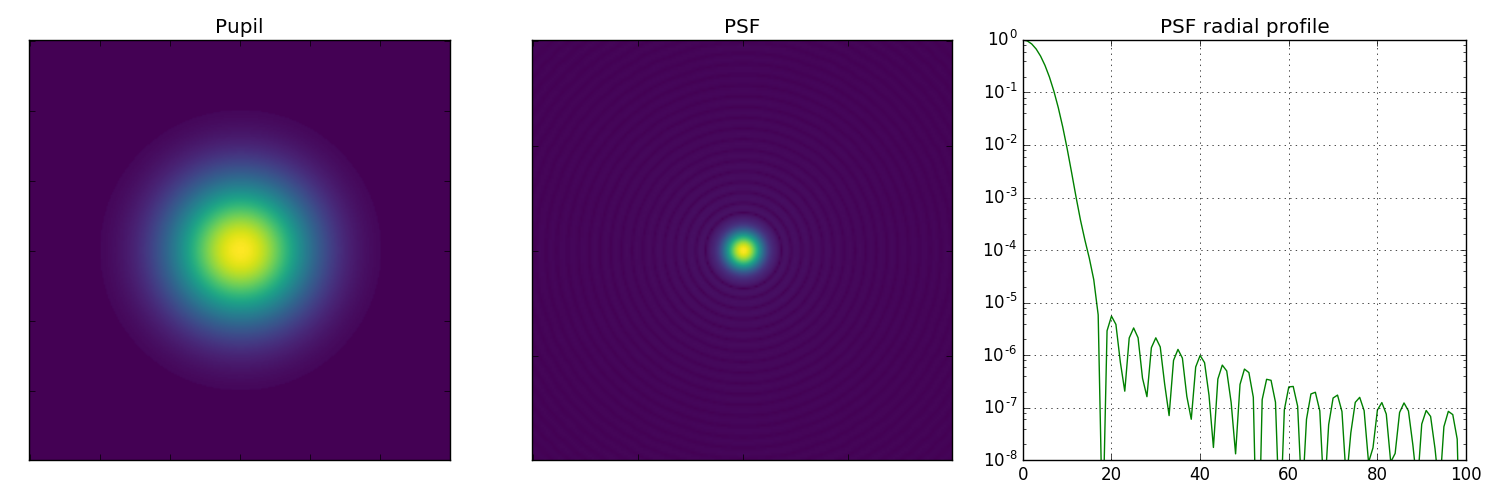}
  \includegraphics[width=\textwidth]{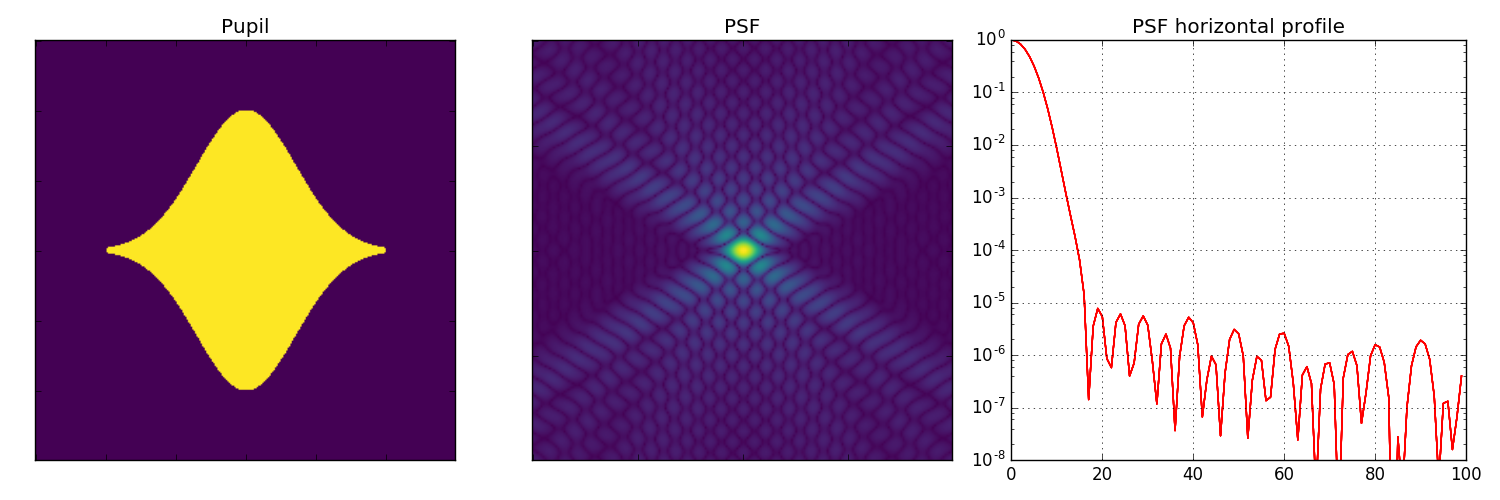}
  \caption{Pupil apodization: the diffraction features of the unobstructed
    circular aperture (top row) are compared to those produced by two
    apodization techniques, using either a variable transmission radial mask
    (middle row) or a binary wave-shaped mask (bottom row).
    The throughput is reduced from 100 \% to $\sim$25 \% for the continuous
    apodization presented in the middle row and to $\sim$56 \% for the binary
    apodization presented in the bottom row. In both cases, the original PSF
    (top row) is replaced by a PSF better suited to the detection of
    high-contrast companions. The PSF profiles visible in the right-hand column
    of panels, using a logarithmic scale, reveal that both apodization
    approaches can produce PSFs with diffraction features with a contrast that
    rapidly drops below 10$^{-5}$ offering a high-contrast detection advantage
    over the full aperture for angular separations as low as three resolution
    elements ($\sim$ 3 $\lambda/D$).}
  \label{f:apod}
\end{figure}

We know that the properties of the PSF of a diffractive aperture are directly
related to the Fourier transform of the illumination of that aperture. The
diffraction rings observed in the PSF of a circular telescope can be attributed
to the sharp transmission edge of the pupil. By tuning the transmission profile
of the aperture, one can expect to be able to alter the PSF and its diffraction
features. This procedure is referred-to as an apodization\footnote{apodization
  litterally refers to the process of removing something's (or someone's!)
  foot} and it can result in a PSF featuring no diffraction rings.  Figure
\ref{f:apod} shows how this apodization effect can be achieved using a fairly
simple shaped-pupil mask placed over the original aperture of the telescope. At
the cost of some throughput (corresponding to the original aperture surface now
covered by the apodization mask), and some angular resolution (the effective
aperture size shrinks because of the mask), the PSF features two symmetric dark
regions at an orientation that can be adjusted by rotating the apodization
mask. The comparison of the PSF profiles represented along the horizontal axis
for both apertures shows that the apodization contributes to reducing the
brightness of the diffraction by more than two orders of magnitude. The energy
previously present in the diffraction rings now contributes to a wider PSF
core, of radius $\sim 3 \lambda/D$. The size of the new PSF core defines what
is now often referred-to as the inner working angle (IWA) of the high-contrast
imaging system.

The solution presented here is by no means optimal: the apodization profile
chosen to produce these figures follow more or less gaussian shapes. In the
litterature, a special class of functions called spheroidal prolates
\cite{2003ApJ...582.1147K, 2003A&A...397.1161S} features properties that make
them ideal for high-contrast imaging, able to deliver theoretical contrast
ratios five orders of magnitude better than those presented in Figure
\ref{f:apod}. Because the properties of the apodized PSF only depend on the
modified pupil shape, the apodized PSF is not more wavelength dependent that
its non-apodized counterpart (see the $1/\lambda$ scaling factor in
Eq. \ref{eq:fourier}), and will be weakly sensitive to pointing
errors. Implemented as it was just described, it however results in throughput
and angular resolution loss as it reduces both the effective collective surface
area and the effective diameter of the aperture.

Apodization can be achieved using as suggested above, with an aperture mask
that suppresses part of the original aperture, or by redistribution of the
light which preserves preserves both throughput and resolution
\cite{2003A&A...404..379G}. The price to pay for one such remapping of the
aperture is a PSF that is no longer position invariant (for which the image -
object convolution relation of Eq. \ref{eq:convol} is no longer valid), at
least until that remapping can be undone \cite{2005ApJ...622..744G}.

To bring its full benefit, the apodization must be adapted to the features of
the aperture \cite{Carlotti2011}: the presence of a central obstruction and
spider vanes in the pupil would render the simple solutions provided in
Fig. \ref{f:apod} useless. The high-performance apodization of real life
telescopes is in practice a complex optimization problem requiring a trade-off
between IWA, overall transmission, and extinction.

\subsection{Coronagraphy}

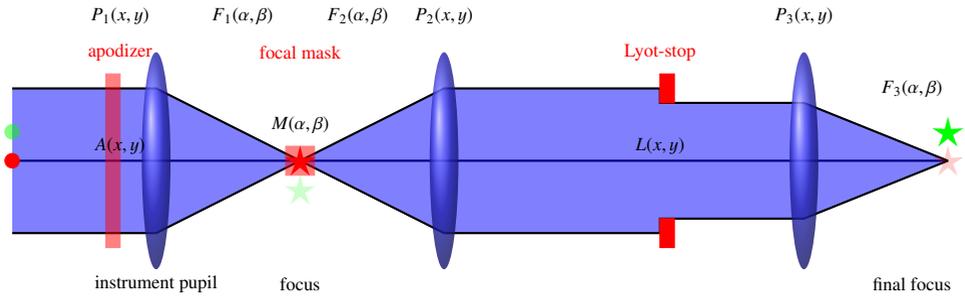
\begin{figure}
\centering
\begin{tikzpicture}[scale=0.95]

  %\draw[black!90] (0,0) grid +(13,4);

  \draw [thick] (0, 2) -- (13, 2); 

  \draw [fill=blue, opacity=0.5] (0,1) -- (2,1) -- (4,2) -- (2,3) -- (0,3);
  \draw [fill=blue, opacity=0.5] (4,2) -- (6,3) -- (9,3) -- (9,2.8) -- (11,2.8) -- 
        (13,2) -- (11,1.2) -- (9,1.2) -- (9,1) -- (6,1) -- cycle;

  \draw [thick] (0, 1) -- (2, 1) -- (6, 3) -- (9,3) -- (9,2.8) -- (11,2.8) -- (13,2); 
  \draw [thick] (0, 3) -- (2, 3) -- (6, 1) -- (9,1) -- (9,1.2) -- (11,1.2) -- (13,2); 
  
  \shade[ball color=blue!80!white,opacity=0.8] 
  (2, 2) ellipse [x radius=0.2cm, y radius=1.5cm];

  \shade[ball color=blue!80!white,opacity=0.8] 
  (6, 2) ellipse [x radius=0.2cm, y radius=1.5cm];

  \shade[ball color=blue!80!white,opacity=0.8] 
  (11, 2) ellipse [x radius=0.2cm, y radius=1.5cm];

  \draw [red, fill=red, opacity=0.5] (1.3,0.8) -- (1.5,0.8) -- (1.5,3.2) -- (1.3,3.2) -- cycle;
  \draw [red, fill=red, opacity=0.5] (3.8,1.8) -- (4.2,1.8) -- (4.2,2.2) -- (3.8,2.2) -- cycle;

  \draw [red, fill=red] (9,1.2) -- (9.2,1.2) -- (9.2,0.8) -- (9,0.8) -- cycle;
  \draw [red, fill=red] (9,2.8) -- (9.2,2.8) -- (9.2,3.2) -- (9,3.2) -- cycle;

  \draw[green, fill=green, opacity=0.5] (0, 2.4) circle (1mm);
  \draw[red, fill=red]                  (0, 2.0) circle (1mm);

  \node [draw=none, scale=2, opacity=0.2] at (4, 1.6) {\color{green}$\star$};
  \node [draw=none, scale=2]              at (4, 2.0) {\color{red}$\star$};
  \node [draw=none, scale=2, opacity=0.2] at (13,2.0) {\color{red}$\star$};
  \node [draw=none, scale=2]              at (13,2.4) {\color{green}$\star$};
  
  % relay optics
  \node at (2,0.3) {\scriptsize{instrument pupil}};
  \node at (4,0.3) {\scriptsize{focus}};
  \node at (12.5,0.3) {\scriptsize{final focus}};

  % coronagraph components
  \node [red] at (1.5,3.5) {\scriptsize{apodizer}};
  \node [red] at (4.0,3.5) {\scriptsize{focal mask}};
  \node [red] at (9.0,3.5) {\scriptsize{Lyot-stop}};

  % symbols
  \node at (1.5,2.2) {\scriptsize{$A(x,y)$}};
  \node at (1.5,4.0) {\scriptsize{$P_1(x,y)$}};
  \node at (3.2,4.0) {\scriptsize{$F_1(\alpha,\beta)$}};
  \node at (4.0,2.5) {\scriptsize{$M(\alpha,\beta)$}};
  \node at (4.8,4.0) {\scriptsize{$F_2(\alpha,\beta)$}};
  \node at (6.0,4.0) {\scriptsize{$P_2(x,y)$}};
  \node at (9.0,2.2) {\scriptsize{$L(x,y)$}};
  \node at (11.0,4.0) {\scriptsize{$P_3(x,y)$}};
  \node at (12.5,3.0) {\scriptsize{$F_3(\alpha,\beta)$}};

\end{tikzpicture}
\caption{Schematic representation of a coronagraph. The light enters from the
  left hand side of the diagram and the final (high-contrast) image plane is on
  the right hand side. The important components of the coronagraph are
  highlighted in red. From left to right: the apodizer, the focal plane and the
  Lyot-stop, placed in a relayed pupil. The relay optics (represented by lenses
  in this diagram) located after the first focus (where the focal plane mask is
  inserted) are required to form a conjugate pupil plane. The final lense is an
  imaging optics that produces the final high-contrast image with the
  appropriate f ratio.}
\label{f:coro}
\end{figure}

Whereas apodization aims at shaping the PSF so as to reduce the impact of the
diffraction rings and spikes, coronagraphy aims at suppressing the light of a
bright source from the focal plane. The technique is slightly more complex than
straightforward apodization as it requires intervention in at least two optical
planes.
Figure \ref{f:coro} provides a schematic representation of the elements
constituing a coronagraph. Three elements are highlighted in red. Going from
left to right, we have: the apodizer that was described earlier, the focal
plane mask, located as its name aptly suggests, in the image plane and the
so-called Lyot-stop, located in a optical plane that is conjugated with the
entrance pupil, after the focal plane mask. While not a part of the original
design of the coronagraph, the benefit of apodization described earlier also
contributes to improving the coronagraph and both techniques are now used
simultaneously \cite{Guerri2011}.

The bulk of the light associated to the on-axis bright source (represented in
Figure \ref{f:coro} by the left-hand side red dot) encounters the focal plane
mask that can either occult it (by absorption or reflection), or dephase it. It
was pointed out earlier that masking out the central part of the PSF alone does
not result in a suppression of the diffraction features outside of this mask.
But if one uses optics to relay the pupil (which is the role of the second lens
in the diagram), the use of a second mask, the so called Lyot-stop, completes
the effect of the focal plane mask and increases the contrast in the final
focal plane. Since it misses the focal plane, the light of an off-axis source
(represented by the green dot located next to the star) is almost entirely
transmitted by the coronagraph and becomes visible in the final focal plane.

\subsection{Coronagraphic formalism}

Using our recently acquired diffraction computation skills, we can complete the
schematic representation of the coronagraph with a formal description of what
is happening at its different stages. Above the different elements represented
in Figure \ref{f:coro}, a few labels are provided that will be used and
referred-to, to describe the complex amplitude of the electric field of
starlight going through the coronagraph.

To better distinguish what is taking place in the pupil from what is happening
in the focal plane, two sets of coordinates are used: linear $(x,y)$ position
coordinates in the aperture and angular $(\alpha, \beta)$ coordinates in the
image. The impact of the three elements of the coronagraph is described by the
apodizing function $A(x,y)$, the focal plane mask function $M(\alpha, \beta)$
and the Lyot-stop function $L(x,y)$. We now know (see Eq. \ref{eq:fourier})
that a Fourier transform $\mathcal{F})$ relates the complex amplitude in the
pupil to the one in the focus: each time the optical system goes back and
forth between pupil and image, a Fourier transform is at work. While cumulating
the effect of consecutive Fourier transform may sound like a terrible idea at
first, it turns out to be fairly simple since:

\begin{equation}
  \mathcal{F}(\mathcal{F}(f(x,y))) = f(-x,-y) =
  \mathcal{R}(f(x,y)),\footnote{The flip of the $(x,y)$ coordinates observed
    after going from pupil to focus and then back to focus, reproduces the
    effect of a lens that produces an inverted image.}
\end{equation}

\noindent
where $\mathcal{R}$ represents the coordinate flip (or reverse) operator.  As
long as we don't land on a detector that records the square modulus of the
complex amplitude (see Eq. \ref{eq:intens}), the elements of the coronagraph
directly interact with the local complex amplitude. This interaction is modeled
by a multiplication by a complex amplitude gain $g$, with a modulus $0 < |g|
\le 1$ (these elements do not amplify the signal) and possibly a phasor
$e^{i\phi}$ term if the component introduces a phase delay $\phi$.
Another nice property of the Fourier transform that helps understand how the
different components affect the final focal plane electric field (and
ultimately the intensity), is the convolution property:

\begin{equation}
\mathcal{F}(f(x,y) \times g(x,y)) = \mathcal{F}(f(x,y)) \otimes \mathcal{F}(g(x,y)),
\end{equation} 

\noindent
that says that the Fourier transform of a product is equal to the convolution
product of individual Fourier transforms. Thus since the impact of an element
of the coronagraph is locally modeled by a multiplication, in the next plane,
it results in a convolution.

\bigskip
With these elements in mind, we can finally describe formally what is taking
place in the coronagraph, using the following sequence of operations:

\begin{itemize}
\item $P_1 = A$ \hfill($A$ and $P_1$ have the same support)
\item $F_1 = \mathcal{F}(P_1)$ \hfill (going to focus $\rightarrow$ Fourier
  transform)
\item $F_2 = M \times F_1$ \hfill (focal plane mask multiplies the complex
  amplitude)
\item $P_2 = \mathcal{F}(F_2)$ \hfill (going back to pupil plane $\rightarrow$
  Fourier transform)
  \subitem $P_2 = \mathcal{F}(M \times F_1)$ \hfill (explicit $F_2$)
  \subitem $P_2 = \mathcal{F}(M) \otimes \mathcal{F}(F_1)$ \hfill (convolution property)
  \subitem $P_2 = \mathcal{F}(M) \otimes \mathcal{R}(P_1)$ \hfill
  ($P_2=$ input pupil convolved by the mask Fourier-transformed)
\item $P_3 = P_2 \times L$ \hfill (the Lyot-stop blocks parts of the pupil)
\item $F_3 = \mathcal{F}(P_3)$ \hfill (going to final focal plane $\rightarrow$
  Fourier Transform)

  \subitem $F_3 = \mathcal{F}(P_2) \otimes \mathcal{F}(L)$ \hfill (convolution property)
  \subitem $F_3 = \mathcal{R}(M \times \mathcal{F}(P_1)) \otimes \mathcal{F}(L)$ \hfill (explicitation of terms)
\item $I = |F_3|^2 $ \hfill (intensity is square modulus of complex amplitude)
\end{itemize}

\begin{sidewaysfigure}
  \vskip11cm
  \includegraphics[width=\textheight]{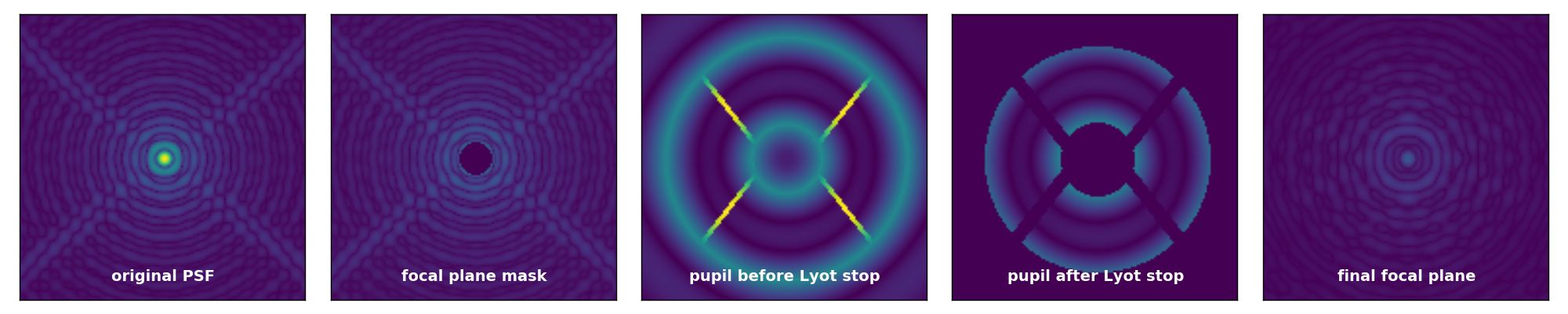}

  \includegraphics[width=\textheight]{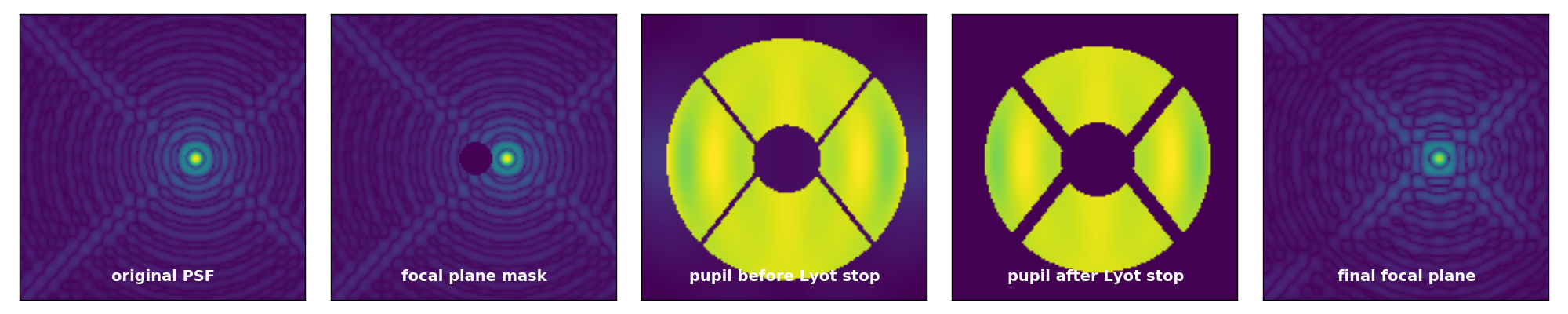}

  \caption{The light going through the differents planes of a coronagraph. The
    top row shows what happens to an on-axis source, whose primary diffraction
    lobe will fall within the region occulted by the focal plane mask (second
    column of panels). The rest of the light is mostly diffracted (in the
    re-imaged pupil plane) near the edges of the aperture (third columns of
    panels) and can be suppressed by an undersized Lyot-stop (fourth row of
    panels). In the final focal plane (fifth column), the light of this on-axis
    star is considerably attenuated. The second row shows what happens to a
    source whose primary diffraction lobe misses the focal plane mask.}
  \label{f:coro_light}
\end{sidewaysfigure}

Figure \ref{f:coro_light} illustrates these different steps by representing the
light of an on-axis (top row) and an off-axis (bottom row) source as it goes
through the different planes of a coronagraph, using no apodization and a
simple focal plane mask occulting the core of the PSF along with its first Airy
ring (radius $\sim2\lambda/D$). The most important property to observe is the
transition from the second to the third column: in the pupil plane that follows
the focal plane mask, the light of the on-axis source is no longer uniformly
distributed but tends to concentrate near the contours (sharp edges) of the
input pupil that features here a central obstruction and spider vanes. To
filter this light that would otherwise find its way back to the final focal
plane, the Lyot-stop masks out these regions, resulting in a slightly
undersized output pupil, with a larger central obstruction and thicker spider
vanes. In the final focal plane, the light of this on-axis source is
considerably attenuated.

The same operations can be applied to the electric field associated to an
off-axis source. The off-axis position will result in a phase slope across the
aperture. It it is sufficiently far off-axis (here $\ge \sim 2\lambda/D$), the
bulk of this electric field off-axis misses the focal plane. In the output
pupil, the light of this source remains mostly uniformly distributed and the
Lyot-stop only induces a reduction of the throughput. In the final focal plane,
the light of this off-axis source is almost integrally transmitted: the
on-axis attenuation combined with a good off-axis transmission results in
images revealing faint structures in the bright star's neighborhood, that would
otherwise remain invisible.

The coronagraph used to produce the images of Figure \ref{f:coro_light} uses an
occulting mask, a configuration known as the classical Lyot-coronagraph
\cite{1932ZA......5...73L} as it replicates (with a smaller occulting mask) the
elements that enabled Bernard Lyot to reveal the corona of the Sun in the early
1930s. A review of the litterature will reveal the existence of a wide variety
of coronagraphs that use different types of masks that can also induce phase
differences \cite{1997PASP..109..815R, 2003A&A...403..369S}, include
subwavelength gratings \cite{2005ApJ...633.1191M} and feature geometries that
split the focus into quadrants \cite{2000PASP..112.1479R,
  2008PASP..120.1112M}. The combination of the coronagraph with an apodizer
\cite{2005ApJ...618L.161S} increases the number of possibilities.

\section{Atmospheric turbulence and Adaptive Optics}
\label{sec:ao}

The purpose of high-contrast imaging devices is to suppress from an image the
on-axis static diffraction signature of an optical system that includes the
telescope, the beam transfer and the instrument optics. The higher the design
performance of the retained solution (often quantified by a level of contrast
at a given separation), the more sensitive that solution ends up being to
changes in the expected system configuration. One important optical element has
however thus far not been taken into consideration: for ground based
observations, the atmosphere ends up being a very important element that can
quickly wreak havoc on the effective coronagraphic performance.
One of the first descriptions of the effect of what we now call atmospheric
turbulence can be found as early as 1704 in Isaac Newton's {\it Opticks}:

\medskip
\noindent
{\it ``If the Theory of making Telescopes could at length be fully brought into
  Practice, yet there would be certain Bounds beyond which Telescopes could not
  perform. For the Air through which we look upon the Stars, is in a perpetual
  Tremor [...] The only Remedy is a most serene and quiet Air, such as may
  perhaps be found on the tops of the highest Mountains above the grosser
  Clouds.''} {\bf Book I, Prop. VIII, Prob. II}.

\begin{figure}
  \hfill
  \includegraphics[width=0.3\textwidth]{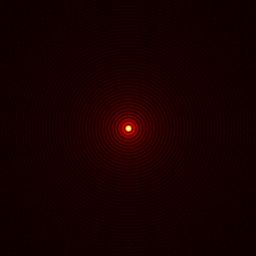}
  \hfill
  \includegraphics[width=0.3\textwidth]{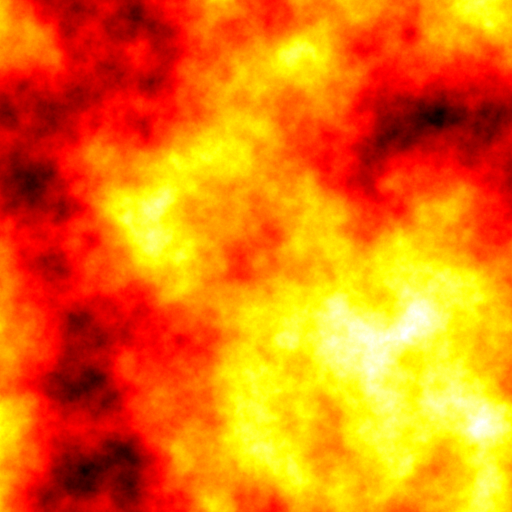}
  \hfill
  \includegraphics[width=0.3\textwidth]{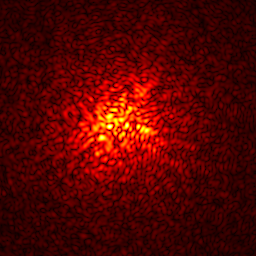}
  \hfill
  \caption{Left: theoretical Airy pattern produced by an unobstructed
    telescope. Middle: example of Kolmogorov phase screen induced by
    atmospheric turbulence. Right: instantaneous seeing-limited point spread
    function experienced when observing through one such atmospheric phase
    screen.}
  \label{f:turbul}
\end{figure}

\medskip
The formation of images through a turbulent atmosphere is a complex process, so
much that atmospheric optics is a research topic on its own. The three
dimensional nature of the atmosphere results in multiple types of degradations:
agitation of the image, spreading of the point spread function due to
high-order wavefront aberrations and scintillation induced by high altitude
turbulence resulting in intensity fluctuations. Figure \ref{f:turbul}
illustrates the typical effect of turbulence for a 1-meter diameter telescope
observing in the visible: the original PSF on the left, with most of the light
($\sim$ 84 \%) concentrated over a $\sim2\lambda/D$ disk is replaced by a
random speckle pattern that extends over a much larger area, suggesting the
existence of smaller diffractive structures (the atmospheric turbulence cells)
\cite{Kolmogorov1941}. The typical dimension $r_0$ of these cells is called the
Fried parameter \cite{Fried1966} and the turbulence characteristic evolution
time $t_0$ depends on the ratio between $r_0$ and the velocity $v$ of the
turbulent layers. A good observing site is characterized by a large $r_0$,
meaning that the turbulence is weak and a large $t_0$, meaning that it moves
slowly. Typical turbulence properties for an average site in the visible
($\lambda$ = 500 nm) are:

\begin{itemize}
\item $r_0 \sim$ 10 cm
\item $v \sim$ 10 m/s
\item $t_0 \sim$ 3 ms
\end{itemize}

What this means is that the high angular resolution potential is no longer just
set by the size of the aperture, but also by the properties of that
turbulence. In the diffraction scenarios discussed so far, the distribution of
complex amplitude for an on-axis source was assumed to be constant across the
diffractive aperture: the wavefront was assumed to be perfectly flat. The
atmospheric turbulence drastically alters this situation and introduces random
phase delays that corrugate the wavefront (see the middle panel of Figure
\ref{f:turbul} for an example of phase screen).

The structure of the wavefront is not entirely random and is driven by
thermodynamics \cite{Kolmogorov1941}. One example of Kolmogorov phase screen is
represented in the middle panel of Fig. \ref{f:turbul}.  The variance between
two parts of the wavefront separated by the distance $\rho$:

\begin{equation}
  D_{\Phi}(\rho) = \left< |\Phi_a(r) - \Phi_a(r+\rho)|^2 \right>_r
\end{equation} 

\noindent
is a 2$^{nd}$ order structure function characterized by one single parameter
$r_0$ introduced earlier as Fried's parameter, so that:

\begin{equation}
  D_{\Phi}(\rho) = 6.88 \left( \frac{|\rho|}{r_0}\right)^{5/3}.
\end{equation} 

The power spectrum of the phase deduced for a Kolmogorov phase screen
\cite{Tatarskii1961, Tatarskii1971}:

\begin{equation}
  W_\Phi(f) = <|\mathcal{F}(\Phi(\rho))|^2> = 0.0228 \: r_0^{-5/3} \: f^{-11/3},
  \label{eq:atmo_plaw}
\end{equation}

\noindent
shows that the distribution of phase follows a power law with a negative
coefficient, which means that the atmosphere introduces more low order
aberrations (associated to a low $f$) such as tip-tilt (pointing), focus,
astigmatism and coma, than high spatial frequencies. The computation of the
diffraction by the aperture (see Eq. \ref{eq:fourier}) is still possible in the
presence of turbulence, but the complex amplitude in the aperture $A(X,Y)$ must
now include the atmospheric-induced phase delay $\Phi$, so that: $A(X,Y) =
e^{i\Phi(X,Y)}$.

The impact of the Kolmogorov phase screen is visible on the right hand side
panel of Figure \ref{f:turbul} that features a short exposure image that keeps
changing with a characteristic time $t_0$.  One can see that while the PSF
spreads out, it is still made of small structures called speckles whose
characteristic size remains of the order of $\lambda/D$, suggesting that some
high-order spatial frequency content can be recovered from the images if one is
able to acquire them with an exposure time of the order of $t_0$. This is the
object of speckle interferometry \cite{1970A&A.....6...85L, Aime2001} which
won't be discussed here.  A long exposure image through turbulence would wash
out these speckles and result in an extended smooth PSF, characterized by a
full width half-max of the order of $\lambda/r_0$.

Under such observing conditions, a high-contrast imaging device like a
coronagraph, originally designed to take out the static component of the
aberration, has very little chance of contributing to a contrast improvement in
the image. The energy associated to the flux of the bright star, previously
concentrated in the central diffraction feature ($\sim$ 84 \%) is now spread
out over a wide number of fainter speckles. The same thing is also happening to
the image of any other source in the field, resulting in an even lower chance
of detecting any faint structure near the bright target. Corrective measures
have to be taken to restore the wavefront entering the coronagraph and make it
as flat as possible again.

\begin{figure}
  \tikzstyle{arrow} = [ultra thick,->,>=stealth]
  \centering
  \begin{tikzpicture}[scale=0.8]

    %\draw[black!90] (0,0) grid +(14,7);
    
    \draw [ultra thick] (0,6) -- (2,6) -- (5, 4.5) -- (8, 4.5) -- (7, 5.5) -- (5, 5.5) -- (2,4) -- (0,4); 
    \draw [fill=blue, opacity=0.5] (0,6) -- (2,6) -- (5, 4.5) -- (8, 4.5) -- (7, 5.5) -- (5, 5.5) -- (2,4) -- (0,4) -- cycle; 
    
    \draw [ultra thick] (7, 5.5) -- (7, 2) -- (7.5,0) -- (8,2) -- (8,4.5);
    \draw [fill=blue, opacity=0.5] (7, 5.5) -- (7, 2) -- (7.5,0) -- (8,2) -- (8,4.5);
    
    \draw [ultra thick] (7,4) -- (9,4) -- (11,3.5) -- (9,3) -- (7,3);
    \draw [fill=blue, opacity=0.5] (7,4) -- (9,4) -- (11,3.5) -- (9,3) -- (7,3) -- cycle;
    
    \draw [fill=blue, opacity=0.5] (6.8, 2.8) -- (8.2,2.8) -- (8.2,4.2) -- (6.8,4.2) -- cycle;
    \draw [ultra thick] (6.8, 4.2) -- (8.2,2.8);
    \draw [ultra thick] (6.8, 2.8) -- (8.2,2.8) -- (8.2,4.2) -- (6.8,4.2) -- cycle;
    
    % cameras
    \draw [red, fill=red] (7,0) -- (8,0) -- (8,-0.2) -- (7,-0.2) -- cycle;
    \draw [red, fill=red] (11,3) -- (11,4) -- (11.2,4) -- (11.2,3) -- cycle;
    
    % DM
    \draw [green, fill=green] (8,4.5) -- (8.5,5) -- (7.5,6) -- (7,5.5) -- cycle;
    \draw [arrow, green, ->] (11.2, 3.5) .. controls (13,4) and (12,8) .. (8,5.5);
    
    % -- wavefronts --
    \draw [ultra thick] (1, 3.8) -- (1.2,4.5) -- (1,5) -- (1.2, 5.5) -- (1,6.2);
    \draw [ultra thick] (5.5, 4.3) -- (5.7,4.7) -- (5.5,5) -- (5.7, 5.3) -- (5.5,5.7);
    %\draw   [ultra thick] (8.5, 2.8) -- (8.7,3.2) -- (8.5,3.5) -- (8.7, 3.8) -- (8.5,4.2);
    %\draw   [ultra thick] (6.8,2.5) -- (7.2,2.3) -- (7.5,2.5) -- (7.8,2.3) -- (8.2,2.5);
    
    \draw  [ultra thick] (8.5, 2.8) -- (8.5,4.2);
    \draw  [ultra thick] (6.8,2.5) -- (8.2,2.5);
    
    % -- lenses --
    \shade[ball color=blue!80!white,opacity=0.8] 
    (2, 5) ellipse [x radius=0.2cm, y radius=1.5cm];
    
    \shade[ball color=blue!80!white,opacity=0.8] 
    (5, 5) ellipse [x radius=0.2cm, y radius=0.7cm];
    
    \shade[ball color=blue!80!white,opacity=0.8] 
    (7.5, 2) ellipse [x radius=0.7cm, y radius=0.2cm];
    
    \shade[ball color=blue!80!white,opacity=0.8] 
    (9, 3.5) ellipse [x radius=0.2cm, y radius=0.7cm];
    
    % legend
    \node at (2,3) {\scriptsize{\bf{telescope}}};
    \node at (7.5,6.5) {\scriptsize{\bf{deformable mirror}}};
    \node at (9,0.5) {\scriptsize{\bf{imaging instrument}}};
    \node at (11,2.5) {\scriptsize{\bf{wavefront sensor}}};
    \node [green] at (10.5,5.8) {\scriptsize{\bf{feedback}}};
    \node [green] at (10.5,5.5) {\scriptsize{\bf{loop}}};
    
    \node at (1.0, 6.6) {\scriptsize{\it{aberrated}}};
    \node at (1.0, 6.3) {\scriptsize{\it{wavefront}}};

    \node at (6.0, 2.7) {\scriptsize{\it{corrected}}};
    \node at (6.0, 2.4) {\scriptsize{\it{wavefront}}};

\end{tikzpicture}
\caption{Schematic representation of the different elements of a closed-loop
  adaptive optics-fed imaging system. Starlight enters from the left-hand side
  of the diagram via the telescope, represented by a single lens. The impact of
  the atmospheric turbulence is represented by the W-shaped wavefront, that
  propagates through the system. Relay optics make it possible to project the
  pupil of the telescope onto a deformable mirror (DM) whose shape can be
  adjusted to compensate the atmospheric effect on the current wavefront before
  it feeds the imaging instrument. To determine the shape the DM should take,
  some light is sampled before the imaging instrument and fed to the wavefront
  sensor (on the right-hand side of the diagram). The analysis of the
  information collected by the wavefront sensor will drive a feedback loop that
  acts on the DM and results in an improved image.}
\label{f:ao}
\end{figure}
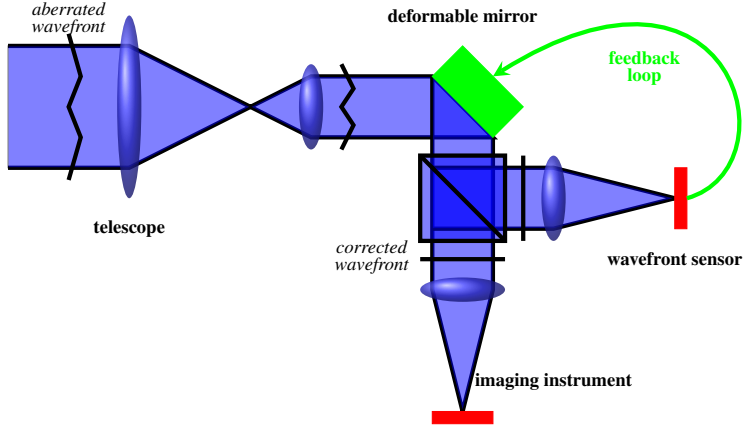

This real-time compensation of the wavefront is the goal of the technique known
as adaptive optics (AO). First described in the 1950s \cite{Babcock1953}, and
deployed by civilian astronomers in the early 1980s \cite{Rousset1999}, AO is
now a tool available at all major ground based observing facilities that
exists in a wide variety of flavors: single or multi-conjugated, involving
natural guide stars (NGS) or artificial (laser) guide stars (LGS). For the
applications discussed here, AO is used in its simplest possible form: NGS -
SCAO. Indeed, because it is focused on a very small field of view (of the order
of one arc-second), high-contrast imaging requires single-conjugated adaptive
optics (SCAO) and its targets, which are nearby stars, are bright enough to
serve as the guide star for the adaptive optics.

An AO system requires two basic functionalities. The first is wavefront
control, that is the ability to act on the wavefront of sources present inside
the field of view, typically (but not only, as we will see later), to flatten
it so as to improve image quality. The second is wavefront sensing, that is the
ability to diagnose what is wrong with the current input wavefront, and to
determine what can be done in order to correct for it. Figure \ref{f:ao}
provides a schematic representation of how these two elements are combined to
make up an AO system.  It should be pointed out that wavefront sensing can take
a wide variety of flavors such as the Shack-Hartmann, the curvature sensor
\cite{1988ApOpt..27.1223R} or the pyramid sensor \cite{Ragazzoni1996}. The
ideal wavefront sensor simultaneously combines good sensitivity, ie. the
ability to operate on faint guide stars; linearity, ie. the ability to run an
unambiguous diagnosis of the wavefront; and a large capture range, ie. the
ability to operate in the presence of large or small wavefront errors. Real
life sensors all seem to be able to only simultaneously gather two of these
qualities at a time \cite{2005ApJ...629..592G}, which means that the choice of
the wavefront sensor will have consequences on the final outcome. This topic
will not be further discussed here and readers interested in this topic are
invited to refer to textbooks dedicated to the topic of adaptive optics
\cite{Roddier2004, Hardy1998}. We however need to take a closer look at the
wavefront control to be able to understand some important features of
AO-corrected images.

\begin{figure}
  \centering
  \includegraphics[width=7cm]{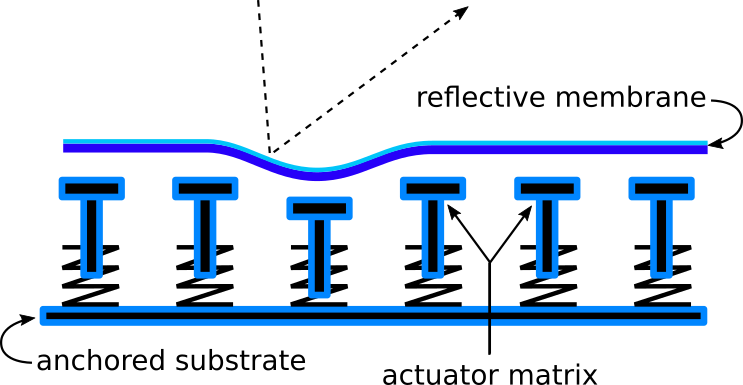}
  \caption{Schematic representation of the layout of a row of actuators pushing
    or pulling on a continuous reflective membrane. The combined effect of the
    motion of all actuators gives the ability to generate complex shapes to the
    deformable surface to compensate the effect of upstream aberrations.}
  \label{f:dm_grid}
\end{figure}

\begin{figure}
  \hfill
  \includegraphics[width=0.49\textwidth]{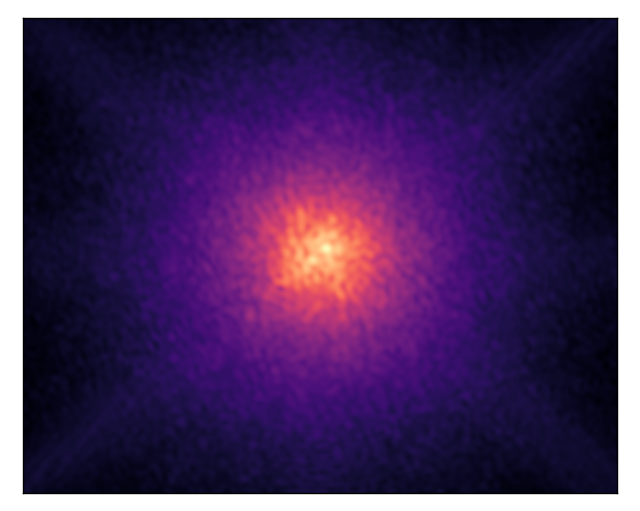}
  \hfill
  \includegraphics[width=0.49\textwidth]{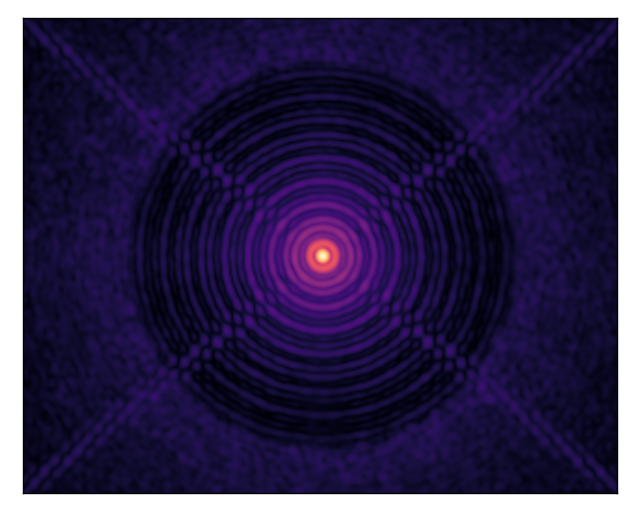}
  \hfill
  \caption{Long exposure acquired through the atmosphere. Left: long exposure
    acquired in the absence of adaptive optics. Unlike the sharp speckle
    pattern presented in Fig. \ref{f:turbul}, the features of this image are
    washed out by the many turbulence realisations forming a wide halo. Right:
    long exposure at the focus of an XAO system. The circular region
    surrounding the now clearly defined and well corrected PSF core at the
    center of the image marks the domain of spatial frequencies that the DM is
    able to compensate.}
  \label{f:longexpo}
\end{figure}

It was shown earlier, that the atmospheric turbulence is characterized by a
power spectrum with a -11/3 power law coefficient (see
Eq. \ref{eq:atmo_plaw}). While the negative sign ensures that less power is
contained in the high-spatial frequencies, there is no limit to how fine the
turbulence structures get in this description: to correct for everything would
require a deformable mirror with an infinite number of active elements, which
is not a realistic solution. In practice, a DM is made of a finite number of
actuators used either to deform a thin reflective membrane or to push and
orient non-deformable mirror segments. For the DMs we are concerned with here,
the actuators are laid out on a regular grid. Figure \ref{f:dm_grid} proposes a
schematic representation for the implementation of a row of actuators. The
total number of actuators distributed across the pupil of the instrument will
determine the finesse of the correction one can expect to produce: the DM will
act like a filter that can attenuate the atmospheric phase screen up until a
cut-off spatial frequency $f_c$ imposed by the number of actuators across
aperture. For the high-contrast imaging application, the wavefront quality
requirement drives the need for a large number of actuators, of the order of a
few thousand for an 8-meter aperture. With such a large number of actuators,
one sometimes talk about extreme adaptive optics or XAO.  The effect of this
cut-off spatial frequency is visible in the right panel of Figure
\ref{f:longexpo}: a clean circular area surrounds the well corrected PSF core
around which one can distinguish the diffraction rings. The DM used to produce
this simulated image features $N=50$ actuators across the aperture, pushing the
correction radius to $r_c = N/2 = 25 \lambda/D$. For an 8-meter aperture
observing in the H-band, this translates into a control region that is $\sim$1
arc second.  Beyond this correction radius, the power contained in the
high-spatial frequency content of the PSF is no-longer compensated and
contributes to the formation of another halo.

\section{Extreme adaptive optics}

The high quality wavefront correction required for high-contrast imaging pushes
for AO systems with a large number of actuators, tightly integrated with the
coronagraph: the integration of the high contrast imaging constraint to the
wavefront control loop marks the specificity of what is now known as extreme
adaptive optics. When deforming the mirror, the distribution of complex
amplitude $A$ across the aperture of the instrument is given by:

\begin{equation}
  A(x,y) = P(x,y) \times e^{i\Phi(x,y)}
  \label{eq:camp}
\end{equation}

\noindent
where $P(x,y)$ describes the shape of the telescope aperture and $\Phi(x,y)$
the distribution of phase from the combined effect of the atmosphere and the
correction by the DM. In the XAO regime, the amplitude of the residual phase is
small enough to justify linearizing the complex amplitude:

\begin{equation}
  A(x,y) \approx P(x,y) \times (1 + i \Phi(x,y)).
  \label{eq:lin_phase}
\end{equation} 

Perfect control of the aberrations would mean $\Phi(x,y) = 0$ over the entire
aperture, leaving only the unity factor resulting in the static diffraction
pattern. This linearized form makes it possible to separate the static and the
dynamic components of the diffraction, respectively corresponding to the real
and imaginary parts of Eq. \ref{eq:lin_phase}. This form can in turn be used to
compute an approximation for the PSF in the low-aberration regime:

\begin{equation}
  \mathrm{PSF}(\alpha, \beta) = || \mathcal{F}(P(x,y)) ||^2 +
  || \mathcal{F}(P(x,y) \times \Phi(x,y)) ||^2.
  \label{eq:low_aber_img}
\end{equation} 

Figure \ref{f:coro_light} showed how the Lyot-coronagraph manages to attenuate
the static diffraction of an on-axis source, but nevertheless leaves a
residual: this coronagraph is not perfect. However, other more recent
coronagraphic solutions like the vortex \cite{2010ApJ...709...53M} and the PIAA
\cite{2003A&A...404..379G} coronagraphs are closer to being able to completely
get rid of this diffraction term \cite{2006ApJS..167...81G}. With such designs,
the post-coronagraphic residuals for the on-axis source are dominated by the
wavefront errors. Coronagraph designs can therefore be benchmarked against the
so-called {\it perfect coronagraph}, a theoretical design producing the
following on-axis coronagraphic image:

\begin{equation}
  \mathrm{I}(\alpha, \beta) = ||\,\mathcal{F}(P(x,y) \times \Phi(x,y)) \,||^2,
  \label{eq:perfect_coro}
\end{equation} 

\begin{figure}
  \centering
  \includegraphics[width=\textwidth]{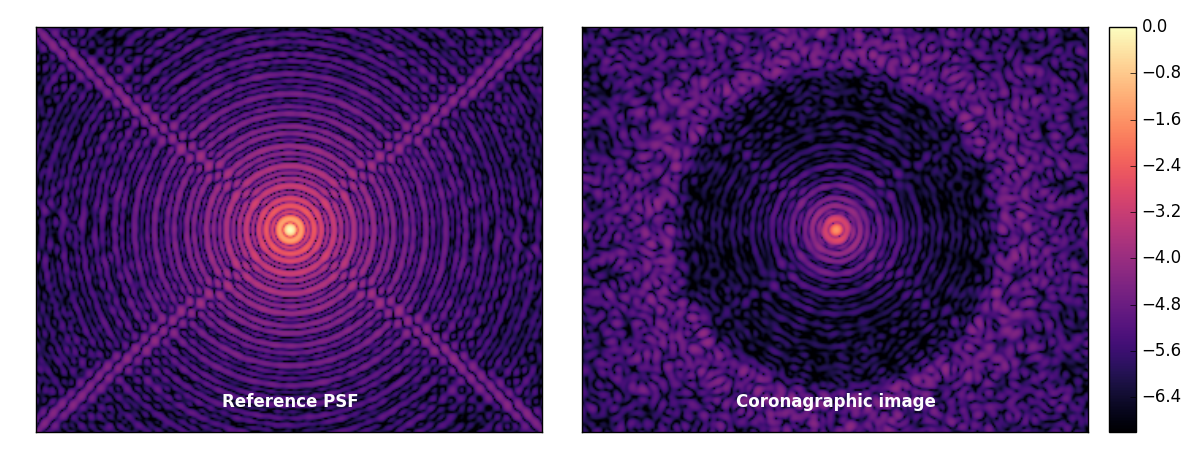}
  \caption{Example of image computation using the perfect coronagraph
    formula. Left panel: a non-coronagraphic PSF, affected by 50 nm RMS
    residual wavefront errors. Right panel: a coronagraphic image computed for
    the same instrument pupil aberration. Both images share the same
    logarithmic stretch and colorbar. }
  \label{f:perfect_coro}
\end{figure} 

\noindent
for which the static diffraction term originally present (see
Eq. \ref{eq:low_aber_img}) has deliberately been removed. In the absence of
aberrations, the perfect coronagraph provides a perfect extinction of the
on-axis source. In the presence of aberrations, the perfect coronagraph leaks
and some residual starlight finds its way to the final focal plane. Figure
\ref{f:perfect_coro} shows one example of application of this perfect
coronagraph formula, for an AO-corrected wavefront residual of 50 nm. The
details of this computation will depend on the statistics of the residual
aberrations. In Eq. \ref{eq:perfect_coro}, we see the phase $\Phi$ squared
appearing as a scaling factor for the wavefront aberration residual light in
the post-coronagraphic focal plane. A close look at the well-corrected area of
the coronagraphic image shown in Fig. \ref{f:perfect_coro} shows that the
coronagraphic leak does look like a scaled-down copy of the original PSF. For a
given RMS level of residuals $\alpha$ (expressed in nanometers) one can
therefore predict a focal contrast improvement factor $c_b$ (for contrast {\it
  boost}) at any place in the focal plane:

\begin{equation}
  c_b = \frac{1}{\sqrt{2}} \biggl(\frac{2\pi\alpha}{\lambda}\biggr)^2.
\end{equation} 

\begin{figure}
  \centering
  \includegraphics[width=0.8\textwidth]{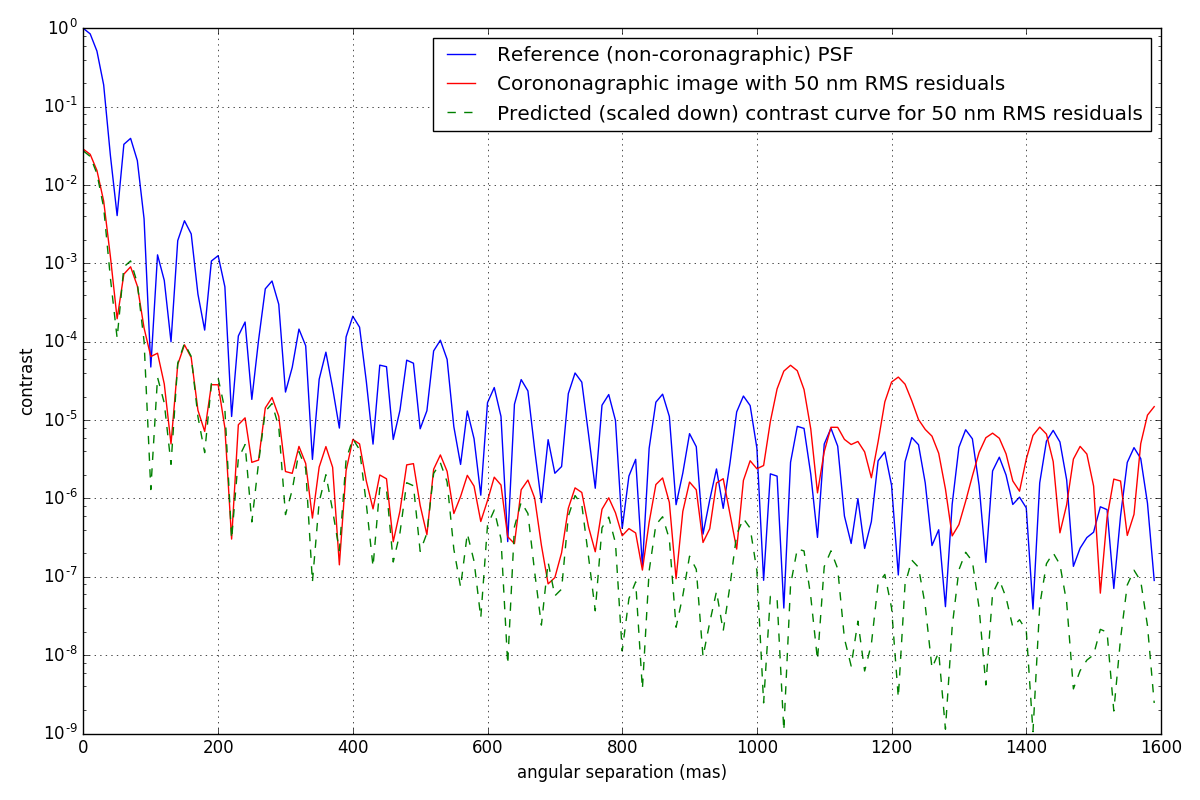}
  \caption{Contrast curves for the perfect coronagraph, in the presence of 50
    nm RMS residual wavefront aberration. Three curves are represented,
    corresponding to radial profiles of the images provided in
    Fig. \ref{f:perfect_coro}: the reference (non-coronagraphic) PSF in blue
    and the corresponding coronagraphic equivalent in red. The green dashed
    line is a copy of the blue curve, scaled down by the factor $c_b$. Over the
    first $\sim500$ mas of the plot, the match between the red and the dashed
    green curves is quite good.}
  \label{f:perfect_coro_curve}
\end{figure}

For the simulated 50 nm RMS residual wavefront error shown in
Fig. \ref{f:perfect_coro}, one therefore expects a $c_b \approx 0.027$ contrast
improvement over the original non-coronagraphic PSF. On
Fig. \ref{f:perfect_coro_curve}, one can verify that over the first $\sim$500
mas of the corrected area, this approximation does match reasonably well the
simulated image.

This model can be further refined \cite{Herscovici-Schiller2017} and used to
derive the statistics of the wave amplitude at each point in the focal plane
\cite{2004ApJ...612L..85A}, and evaluate the relevance of high-contrast devices
in general. What this kind of study shows is that it is not useful to design a
coronagraph that attenuates the on-axis PSF of a bright star further than the
amount of residual speckles expected for a given amount of residual wavefront
aberrations. It is the performance of the AO that will drive how far a
coronagraph can help you go, and we will now look into ways the AO performance
can unfortunately throw you off target.

\section{Calibration of biases}

So far, the description has been assuming that the AO system was doing the
right thing: to flatten the wavefront so as to help the coronagraph effectively
erase the on-axis static component of the diffraction of a bright star. This
turns out to be a somewhat naive assumption, due to a simple but fundamental
limitation. The wavefront sensor (refer back to Fig. \ref{f:ao}) can indeed
only sense the aberrations introduced by the instrument optics all the way down
to the optics that splits the light between the sensor and the downstream
instrument. The AO is therefore understandably oblivious to anything affecting
the light on the instrument path after this split which results in practice in
a non-common path error. Much care is obviously taken while initially setting
up instruments, to minimize any non-common path error however, ground based
instruments are not immune to minute temperature changes and mechanical
flexures. Given the very strong dependence of the coronagraphic rejection on
input wavefront quality, the least amount of non-common path error will
considerably reduce the discovery potential of any high-contrast imaging
instrument.

\begin{figure}
  \centering
  \includegraphics[width=.5\textwidth]{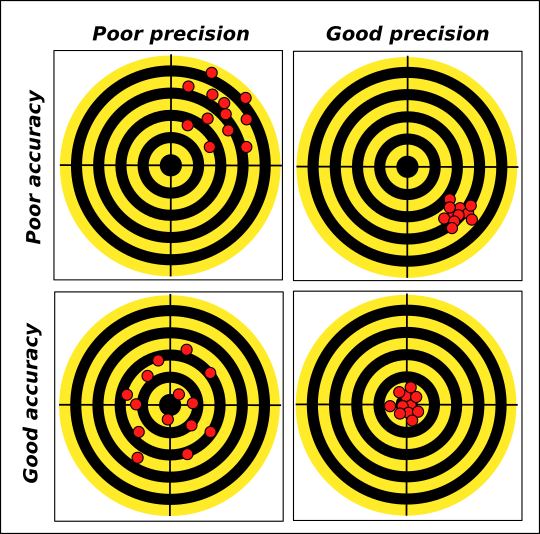}
  \caption{Precision and accuracy are two important statistical concepts that
    characterize all measurements. As we improve the quality of our
    instrumentation, usually reducing the precision of the measurements they
    produce, the impact of systematic bias affecting the accuracy of our
    conclusions becomes an essential element to take into consideration. Much
    energy should be spent on understanding the origin of biases and figuring
    out how to calibrate them out.}
  \label{f:bias}
\end{figure}

These techniques are victims of their own success: before the generalized use
of AO, the overall quality of astronomical images produced by a well built
instrument was dominated by random atmospheric induced errors. The progressive
deployment of AO and the improvement of its performance has reduced the
contribution of random errors, resulting in an {\it improved precision}. But
when the amplitude of random errors is reduced, the effect of small but
systematic biases affecting our {\it accuracy} become more apparent (see
Fig. \ref{f:bias}). XAO-fed high contrast imaging and long baseline
interferometry make it possible to enter the realm of very high-precision
observations: the search and compensation of instrumental biases is becoming
more important than ever.

Despite the very high quality wavefront control (50 nm RMS only) used to
produce the simulated coronagraphic image in Fig. \ref{f:perfect_coro}, the
control region features a large number of speckles amongst which a faint
genuine companion to the bright star could hide. If induced by random AO
residual errors, the amplitude noise induced by these structures can be reduced
simply by accumulating enough data. However if some static or quasi-static
speckle structures induced by an aberration that is not seen by the wavefront
sensor persist over long time-scales, the detection of faint planetary
companions is compromised. We are going back to the important question
highlighted in the introduction of this paper: any speckle-like feature present
in the image can either be a diffraction induced artefact or a genuine
structure of the target being observed.

The non-common path error turns out to be one of the dominant limitations of
high-contrast imaging instruments as faint systematic structures are reported
to survive in images over timescales stretching as far as $\sim$1 hour. The
simplest calibration procedure employed in astronomy consists in using images
acquired on a reference object of known characteristics (ideally a featureless
single star), observed under conditions as identical as possible to those used
for the object of interest, and to subtract the calibration image from the
image of the target of interest. Using the shooting analogy used in
Fig. \ref{f:bias}, we would use a series of shots aiming at the center of the
target to figure out how off our aim really is, before going for the big
game. This strategy is common for the interpretation of AO-corrected images as
well as for optical interferometry for which it is often advised to observe
more calibrators than targets of interest.

\begin{figure}
\includegraphics[width=\textwidth]{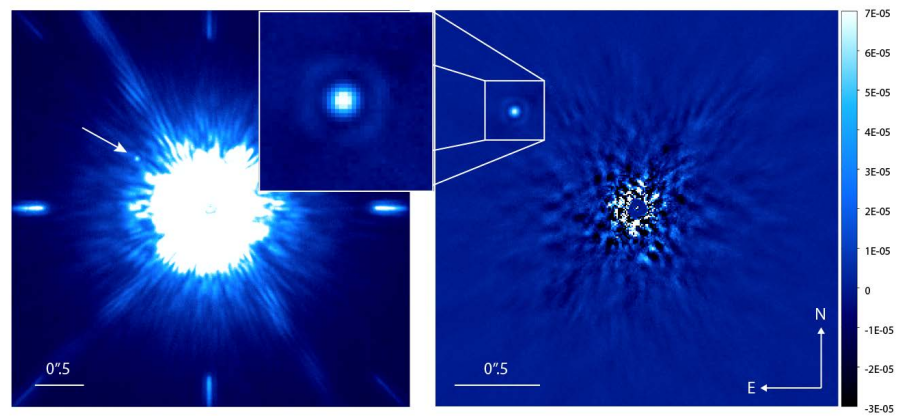}
\caption{Illustration of the impact of post-processing (from \cite{K2018}) on
  high-contrast observations with a vortex-coronagraph by the SCExAO instrument
  \cite{2015PASP..127..890J}. The left hand side panel features shows a raw 15
  second exposure of $\kappa$-{\it And} and the right hand side panel shows the
  impact of the ADI quasi-static speckle pattern subtraction. The highlighted
  companion, that could be mistaken for a speckle visible in the raw image, is
  clearly above the noise level after the post processing has been applied.}
\label{f:post}
\end{figure}

While reasonably effective the calibrating potential of this method remains
limited: it is difficult to guarantee that observing conditions are indeed
strictly identical when going from a science target to a calibrator. The
repointing of the telescope, the small differences in elevation, spectral type
and evolutions of the atmospheric seeing will eventually translate into biases
of their own.  For high-contrast imaging, other approaches are available that
do not require to alternate observations on a target interest with those of one
or more calibration stars: it is indeed possible to take advantage of the field
rotation experienced when an instrument is installed at the focus of an
alt-azimuthal telescope. When following a target as it transits across the
local meridian with one such instrument, the target appears to rotate while the
residual diffraction induced by the instrument corrected by the AO but still
affected by the non-common path error, remains stable. The relative rotation
between the observed scene and the residual diffraction pattern can be used to
distinguish spurious diffraction features from genuine structures in a series
of images. This approach is referred-to as angular differential imaging or ADI
\cite{2006ApJ...641..556M} and has led to a wide variety of algorithms such as
LOCI \cite{2007ApJ...660..770L}. Any type of observation that include some form
of diversity such simultaneous imaging in two spectral bands (spectral
differential imaging or SDI) or in two polarization states (polarized
differential imaging or PDI), theoretically makes it possible to calibrate out
systematic effects that bias observations. One must however remain attentive to
the implementation details, as these techniques end up relying on further
splits of the light path, which can become a source of non-common path. To
account for all possible biases during such observations requires a multiple
tier calibration procedure, that includes the ability to swap light paths, an
example of which is given in \cite{Norris2015}.

More recent approaches give a new spin to the idea of using calibrators, by
relying on the principal component analysis of a library of reference PSFs
\cite{Soummer2012}, which provides performance comparable to ADI-inspired
approaches. Finally, it is also possible to take the information contained in
AO-corrected images (albeit not coronagraphic ones), and to project it onto a
sub-space (called the Kernel) that filters out aberrations
\cite{2010ApJ...724..464M}. The approach is reminiscent of the closure-phase
technique used in interferometry
\cite{1958MNRAS.118..276J,1986Natur.320..595B}, but is now applicable to
AO-corrected images \cite{2016MNRAS.455.1647P}, and is particularly relevant
for detection near the diffraction limit (around $\sim$1-2
$\lambda/D$). Regardless of the algorithmic details at work, Figure
\ref{f:post} shows the impact the post-processing has on high-contrast imaging
by comparing a single raw coronagraphic image to the result of post-processing
of a 10-minute series of images: the impact of the speckle subtraction is
spectacular and sometimes contribute as much if not more than the coronagraph
itself.

\section{Focal-plane based wavefront control?}

Unless high-contrast imaging solutions that are intrinsically robust to weak
amounts of aberrations do emerge, better calibration stategies must be employed
if a performance improvement is desired. The importance for a good calibration
of systematic effects in coronagraphic images will grow as the quality of the
upstream AO correction keeps on improving. One needs to find, at the level of
the focal plane, a discrimination criterion that will make it possible to
distinguish a genuine struture in the focal plane from a spurious diffraction
induced speckle. The introduction to this paper already hinted at one
possibility, relying on the ability to measure the degree of coherence of the
structure in question. Section \ref{sec:coherence} introduced the idea of
coherence as the ability of light to interfere. Given the two important
coherence properties of astronomical sources: the fact that the light of an
unresolved point source is perfectly coherent while the light of distinct point
sources is incoherent, can be used to discriminate speckles from planets in an
image.

Deliberate modulation of the starlight synchronized with acquisitions by the
focal plane camera form the basis for an ideal coherence test that will
discriminate the true nature of the high-contrast features present in an
image. The deformable mirror can indeed be used to send additional light atop
of whatever is currently in the focal plane and the camera can be used to
diagnose the degree of coherence of the light recorded in the live image.

\begin{figure}
  \centering
  \includegraphics[width=0.9\textwidth]{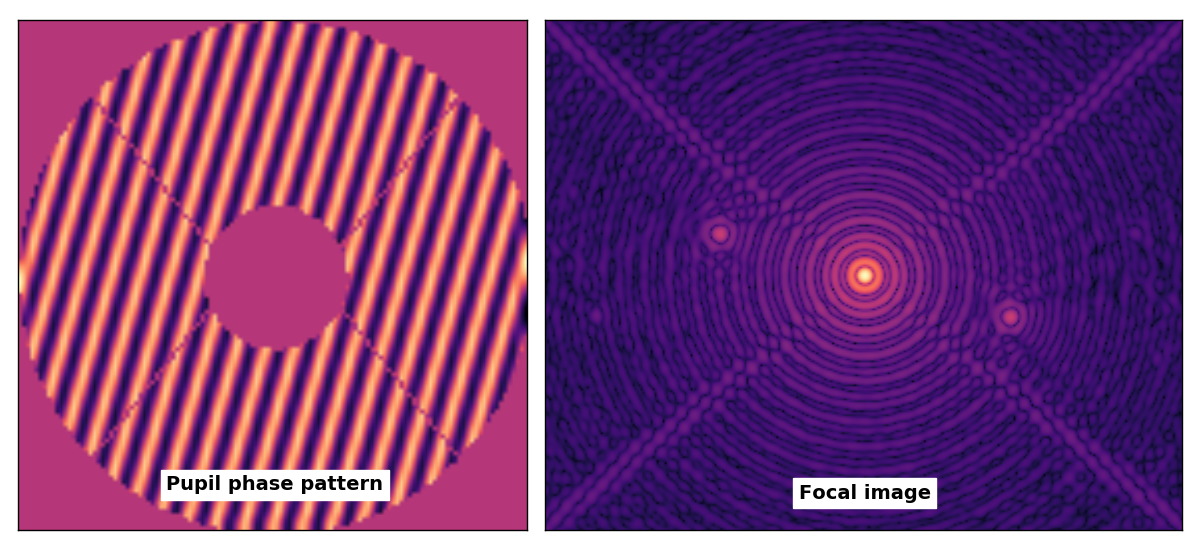}
  \caption{Left: sinusoidal modulation of the wavefront across the instrument
    pupil by the high-order deformable mirror. Right: the resulting point
    spread function. The original on-axis diffraction pattern is flanked by two
    high-contrast replicas at a distance of $\sim$18 $\lambda/D$, given by the
    number of cycles across the aperture.}
  \label{f:replicas}
\end{figure}

The grid structure of the DM actuators (see Figure \ref{f:dm_grid}) used in XAO
systems makes them particularly suited to the generation of sinusoidal
modulation patterns. When one such modulation is applied, the DM behaves like a
diffraction pattern and displaces some of the starlight that would otherwise be
transmitted on-axis (and possibly attenuated by the coronagraph) at a distance
that is proportional to the number of cycles across the aperture.

With $N$ actuators across one aperture diameter, the highest spatial frequency
one can reach corresponds to a state where every other actuator is pushed up
with the others pushed down: the sinusoidal wave thus generated contains $n_c =
N/2$ cycles across the aperture. This is what sets the cut-off spatial
frequency of AO introduced in Section \ref{sec:ao}. A deformation $\Delta$ of
respectively $k_x$ and $k_y$ cycles (both $\le n_c$) along the $x$ and $y$
directions of the image is equal to:

\begin{equation}
  \Delta(x,y) = a \times \sin (2 \pi (k_x x + k_y y) / D + \phi)
\end{equation} 

\noindent
where $a$ is the amplitude of the modulation (typically expressed in microns or
nanometers) and $\phi$ the phase of that modulation. For a small modulation
amplitude $\alpha$, the complex amplitude $A(x,y)$ induced by this deformation
can be linearized (like for Eq. \ref{eq:lin_phase})\footnote{The global scaling
  factor is here $4\pi/\lambda$ and not $2\pi/\lambda$ as one might have
  expected. This $\times2$ factor is there to take into account the fact that
  we are dealing with a reflection off a mirror: a $\Delta$ mechanical
  deformation of the surface induces a $2\Delta$ deformation of the wavefront.
}:

\begin{eqnarray}
  A(x,y) &=& P(x,y) \times \exp{\biggl( i4\pi/\lambda \times
    \Delta(x,y)\biggr)} \\ &\approx& P(x,y) \times \biggl(1 + i4\pi a/\lambda
  \times \sin \bigl(2 \pi (k_x x + k_y y) / D + \phi\bigr)\biggr)
  \label{eq:modul}
\end{eqnarray}

We know that a Fourier transform relates the distribution of complex amplitude
in the pupil to that in the focal plane, and can relate values of $a$ and
$\phi$ to the properties of speckles in the focal plane. If one knows the
Fourier transform of the sine function:

\begin{equation}
  \mathcal{F}\bigl(\sin{(2\pi k x)}\bigr) = \frac{1}{2i} \bigl( \delta(u-k) - \delta(u+k) \bigr),
\end{equation}

\noindent
where $\delta(u)$ is the Dirac distribution, then the Fourier transform of
Eq. \ref{eq:modul} will write as:

\begin{equation}
  \hat{A}(\alpha, \beta) = \hat{P}(\alpha, \beta) \otimes \biggl(
  \delta(\alpha, \beta) + \frac{2\pi a}{\lambda} \biggl(e^{i\Phi}
  \delta(\alpha-k_x, \beta-k_y) - e^{-i\Phi} \delta(\alpha+k_x,
  \beta+k_y)\biggr) \biggr),
  \label{eq:sat_spkls}
\end{equation}

\noindent
using angular coordinates $\alpha$ and $\beta$ expressed in units of
$\lambda/D$.  A detector located in the focal plane will record the square
modulus of this expression. Each $\delta$ function (convolved by the Fourier
transform of the aperture $\hat{P}$) marks the location of one
PSF. Eq. \ref{eq:sat_spkls} therefore allows you to predict that the focal
plane will feature two replicas of the original on-axis PSF, at positions given
by the number of cycles $k_x$ and $k_y$. The two replicas are characterized by
reference complex amplitudes: $\nu_1 = (2\pi a/\lambda) \times e^{i\Phi}$ and
$\nu_2 = (2\pi a/\lambda) \times e^{i(\pi - \Phi)}$. An example of image
showing this is shown in Fig. \ref{f:replicas} for $\sim$18 cycles, resulting
in a pair of symmetrical replicas of the on-axis PSF at $\sim$18 $\lambda/D$.
Whereas the number of cycles imposes the location of the replicas, the
amplitude of the modulation $a$ sets the contrast relative to the original PSF,
which is given by:

\begin{equation}
c = (2\pi a / \lambda)^2.
\end{equation} 

Plugging in a sinusoidal phase modulation of amplitude $a=50$ nm therefore
produces in the H-band ($\lambda \approx $1.6 $\mu$m) a pair of replicas of
contrast $c \approx 4 \times 10^{-2}$ which may seem surprisingly bright to the
reader. Figure \ref{f:perfect_coro_curve} indeed presented for the perfect
coronagraph in the presence of a similar level of RMS error, considerably more
favorable contrasts. The difference between the two scenarios lies in the
structure of the residual phase noise: near random in the case presented in
Fig. \ref{f:perfect_coro_curve} which distributes the total amount of light
associated to the RMS over the control region, or highly structured in the
sinusoidal modulation scenario, that focuses the diffracted light onto two
specific locations. In addition to the residual RMS given by an AO or XAO
system, it is therefore also important to understand how the residuals are
distributed across the aperture.

\medskip

So far, we've accounted for the number of cycles $k_{x,y}$ and the modulation
amplitude $a$ but not for the $\Phi$ and $\pi-\phi$ phase of the replicas
remains: if one were to ignore the convolution operation by $\hat{P}$, taking
the square modulus of second term of Eq. \ref{eq:sat_spkls} would make those
phase terms disappear as the $e^{i\Phi}$ factor disappears, suggesting that the
phase of the replicas does not matter. The convolution will however bring
diffracted light over the area covered by the replica. We have starlight
landing atop of starlight: the two contributions will interfere with one
another. If the light of an incoherent source is present (ie. a planet or one
local disk structure), then the added light will not interfere with this
structure: the two intensitise will simply add. Depending on the phase
difference between the replica and the speckles or diffraction features already
present in the focal plane, the interference can either be constructive or
destructive, which leads to an interesting prospect: the possibility of
improving the raw contrast of images by tweaking the shape of the deformable
mirror. Earlier, it was pointed out that the job of AO feeding a classical
imaging system is to flatten the wavefront so as to improve the overall image
quality: for a high-contrast imager, the optimal strategy is no longer to
flatten the wavefront but to improve the contrast in the focal plane, which can
drive the DM to shape the DM quite far from flat.  This idea was first
envisioned for space \cite{1995PASP..107..386M} and has led to a series of
sophisticated algorithms such as speckle-nulling \cite{2006ApJ...638..488B},
electric field conjugation \cite{2006JOSAA..23.1063G} or stroke minimization
\cite{2009ApOpt..48.6296P}. If the wavefront sensing systems used in modern AO
systems tolerate the idea of being driven away from a flat reference wavefront,
then this approach becomes implementable on ground based high-contrast imaging
instruments.

To produce a fully destructive interference that would result in a local
reduction of the local intensity in the image, the complex amplitude of the
added replica must match that of the already present structure, with the same
amplitude but with opposite phase.

In practice, one does not direcly measure the contrast of speckles: one will
primarily access to a local intensity $I_0$ which we know will be proportional
to the square modulus of the speckle complex amplitude:

\begin{equation}
I_0 = \gamma \times || a_0 e^{i\Phi_0}||^2 = \gamma \times || a_0 ||^2,
\end{equation} 

\noindent
where the proportionality constant $\gamma$ will depend on the brightness of
the target and the exposure time and will therefore have to be regularly
estimated and controlled. While the intensity associated to speckle gives us a
proxy for its amplitude $a_0$, its phase $\Phi_0$ remains unknown. To determine
it, one can use a probing approach, which consists in following the evolution
of the local intensity as the speckle interferes with a probe speckle of known,
stable amplitude $a$ and variable phase $\Phi$. The intensity of the coherent
sum of these two complex amplitudes is the result of a classical two-wave
interference equation:

\begin{eqnarray}
  I(\Phi) &=& \gamma \times \bigg|\bigg|  a_0 e^{i\Phi_0} + a e^{i\Phi} \bigg|\bigg|^2 \\
  &=& \gamma \times \biggl(  a_0^2 + a^2 + 2 a_0 \: a  \cos{(\Phi_0 - \Phi)}
  \biggr).
  \label{eq:spk_mod}
\end{eqnarray}

\begin{figure}
  \centering
  \includegraphics[width=0.8\textwidth]{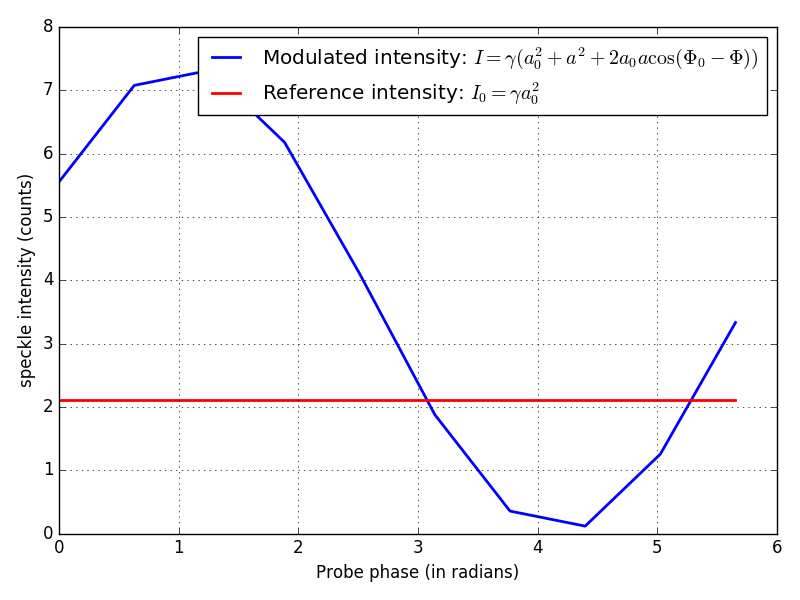}
  \caption{Modulation of the speckle intensity (expressed in units of detector
    counts) as a function of speckle probe phase. The red curve marks the
    intensity of the original speckle for which the phase is unknown. The blue
    curve is the result of the addition of a probe speckle of variable phase
    $\Phi$. The observed modulation confirms that the speckle is a coherent
    (ie. contains starlight) structure: one phase of the added speckle ( $\Phi
    \sim$ 4.5 radians) attenuates the local intensity, resulting in an improved
    local raw contrast.}
  \label{f:spk_mod}
\end{figure}

This interference function is evaluated for an adaptable number of probes
with a phase $\Phi$ uniformly sampled between 0 and 2$\pi$ radians and at least
three distinct values of $\Phi$ are required to constrain the values of $a_0$
and $\Phi_0$. In practice, a finer sampling minimizes the sensitivity to high
temporal frequency phase noise (overall jitter and AO dynamic residuals). An
example of modulation is represented in Figure \ref{f:spk_mod}. It compares the
original intensity level marked by the horizontal red line to the blue curve
recording the evolution of the phase modulation. When the probe is in phase
with the original speckle ($\Phi = \Phi_0$), the local intensity is
quadrupled. When the probe is in phase opposition with the original speckle
($\Phi = \Phi_0 + \pi$), the local intensity can be brought to zero.

With four probes, with phases 0, $\pi/2$, $\pi$ and $3\pi/2$, an analytical
solution exists to directly measure the complex amplitude of the speckle: this
is the so-called ABCD-method. A more general solution is however possible, that
is compatible with an arbitrary number $N$ of phases (with $N \ge 3$). It boils
down to a parametric model fit of the modulation curve. In addition to
$\mathbf{I}_S$ the vector of $N$ intensities recorded during the probing
sequence, one can precompute a separate vector $\mathbf{W}$ that contains
the consecutive powers of the $N^{th}$ root of unity $w_N=e^{i2\pi/N}$. The
value of the phase $\Phi_0$ is directly given by the argument of the dot
product between these two vectors:

\begin{equation}
  \Phi_0 = \mathrm{arg}\biggl(\mathbf{I}_S^\top \cdot \mathbf{W} \biggr),
\end{equation} 

\noindent
while the visibility modulus $\Gamma$ ($0 < \Gamma < 1$) characterizing the
modulation described by Eq. \ref{f:spk_mod} is related to the modulus of the
dot product:

\begin{equation}
  \Gamma = \frac{2 a \: a_0}{a^2 + a_0^2} = \frac{2}{N} \big|| I_S^\top \cdot
  W \big||.
  \label{eq:spk_mod_gamma}
\end{equation}

The amplitude $a_0$ of the original speckle is one of the two roots of the
following quadratic equation:

\begin{equation}
  a_0^2 - \frac{2 a}{\Gamma} a_0 + a^2 = 0,
\end{equation}

\noindent
which are given by:

\begin{equation}
  a_0 = \frac{a}{\Gamma} \bigl( 1 \pm \sqrt{1 - \Gamma^2}\bigr).
  \label{eq:spk_solutions}
\end{equation}

The amplitude $a$ of the probe is selected to be as close as possible to the
amplitude of the speckle $a_0$, so as to maximize the visibility modulus
$\Gamma$, which results in an improved sensitivity to the properties of the
speckle. Because we can't afford to make a mistake that will amplify the
speckle present if we pick the wrong amplitude, one solution is to
systematically buff up the probe (for instance by 5 \%): some sensitivity is
lost but we can be sure that the solution (from Eq. \ref{eq:spk_solutions})
with the minus sign will always be the right one.

\begin{figure}
\includegraphics[width=\textwidth]{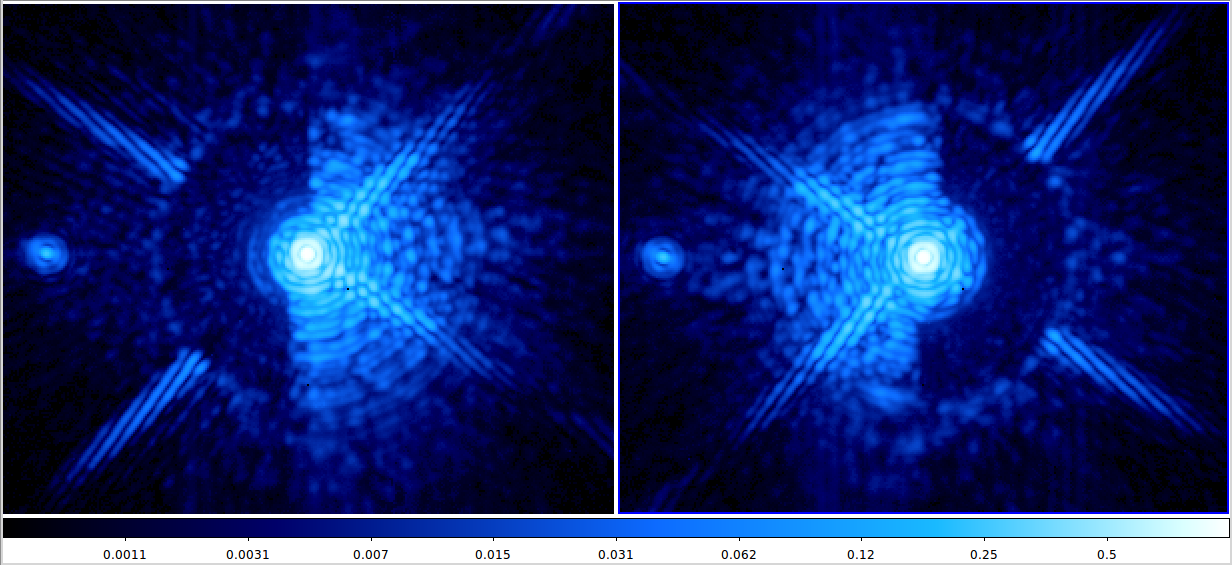}
\caption{Result of two speckle nulling experiments carried out on the SCExAO
  instrument on an internal calibration source without a coronagraph. For each
  image, one can see a high-contrast region was created by the speckle nulling
  loop on one side of the field of view only. Although it is less obvious on
  these images, on the opposite side, the speckles and diffraction features of
  the PSF are amplified as a result of the speckle nulling.}
\label{f:scexao_cal}
\end{figure}

Using this algorithm, it is possible to create a closed-loop focal-plane based
wavefront control loop that modifies the reference position of the DM to create
a higher contrast area within the control region. Note that while the
description of this technique looked at a single speckle, the algorithm can be
multiplexed and simultaneously probe dozens or even hundreds of speckles (the
exact number depends on the number of actuators available), making it much more
efficient. Note that instead of a temporal modulation of the coronagraphic
speckle field inside the control region, spatial modulation is possible: the
self-coherent camera (SCC) \cite{Galicher2008} relies on this idea. Instead of
acquiring a sequence of images before applying a correction, the focal plane
must be oversampled while a reference beam of starlight is uniformly projected
over the control region: the speckles feature fringes that directly encode the
complex amplitude properties of the speckles.

Speckles in the focal plane have two origins: they can either be induced by
pupil phase aberrations, or can be induced by the geometry of the aperture
itself if the coronagraph is absent or if it is imperfect: one can always think
of diffraction rings and spikes as being made up of a coherent sum of
individual speckles: we can refer to these as induced by pupil amplitude
aberrations. It sounds fair game to try and compensate for phase aberration
induced speckles by phase modulations, but what of pupil amplitude induced
ones? By bending the wavefront, the DM can only redistribute the energy in the
focal plane and not make it disappear: when it corrects phase induced speckles,
the energy associated to these speckles gets injected back into the original
PSF; when it corrects an amplitude induced speckle on one side of the field, it
amplifies the speckle on the opposite side. Figure \ref{f:scexao_cal} shows the
result of two speckle nulling experiments done on the SCExAO instrument in a
non-coronagraphic mode. The high-contrast region created by the successive
attenuation of speckles covers half of the control region. The appropriate DM
modulation produces a high-contrast region in the focal plane, an effect that
is similar to what apodization achieves (see Sec. \ref{sec:apod}). The use of
static apodizing phase plates achieving a similar effect \cite{Kenworthy2007}
has been successfully exploited on-sky. These phase plates benefit from
advantages that render them fairly achromatic \cite{Otten2014}. But like all
static high-contrast imaging devices, they are not immune to biases and a focal
plane feedback loop remains essential.

\begin{figure}
\hfill
\includegraphics[width=0.4\textwidth]{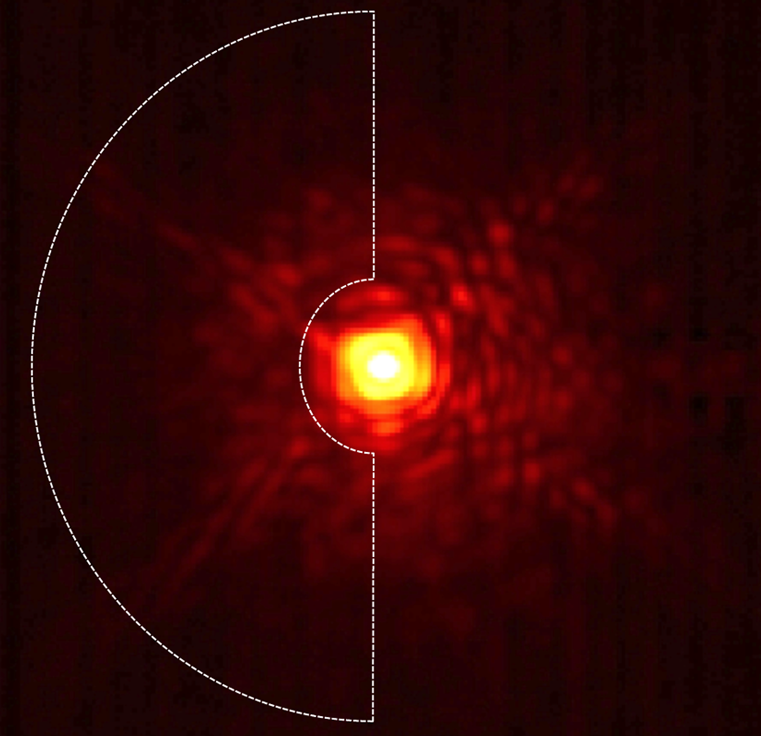}
\label{f:spk_nul_sky}
\hfill
\includegraphics[width=0.4\textwidth]{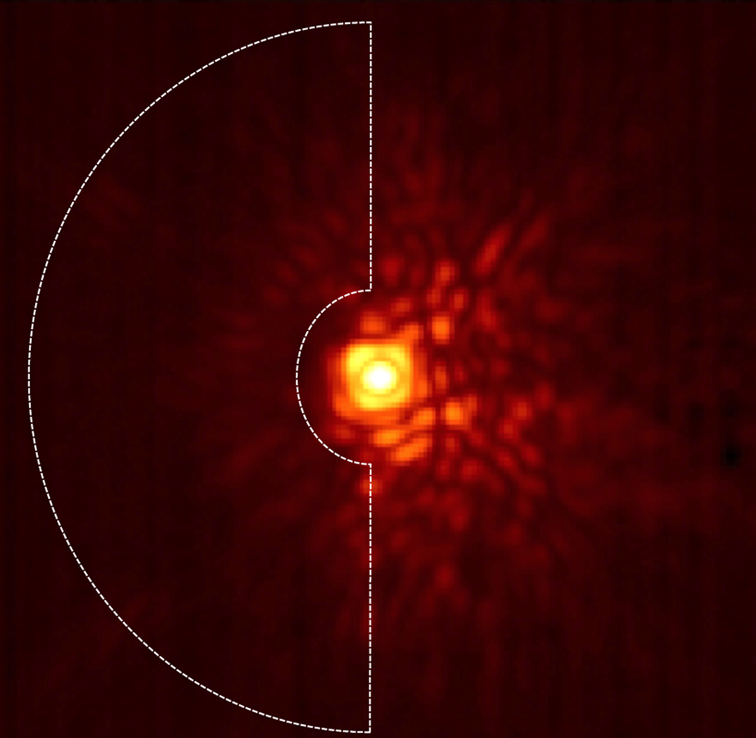}
\hfill
\caption{Example of on-sky speckle nulling experiment with the SCExAO
  instrument without a coronagraph (from \cite{2015PASP..127..890J}). Left: the
  starting point of the algorithm, after the initial lock of the upstream AO
  system. Right: after a few minutes of speckle nulling closed loop operation,
  the left hand side of the field is effectively darker. The D-shaped control
  region over which the loop is operating is highlighted.}
\label{f:scexao_sky}
\end{figure}

If the same deformable mirror is simultaneously driven by the upstream AO that
tends to flatten the wavefront and this focal plane based control that
deliberately pushes it away from the flat, to improve the raw contrast in the
image, conflicts may occur. One (easy but expensive!) solution may be, in
future implementations, to rely on two distinct mirrors for the AO and the
high-contrast. The other (cheaper but more difficult) requires the upstream AO
system to agree with the idea of stabilizing the wavefront away from flat. This
is the object of ongoing work: an exemple of partial speckle nulling correction
obtained on sky also with the SCExAO instrument is presented in Figure
\ref{f:scexao_sky}.

\section{Conclusion}

One of the goals of this lecture was to highlight the formalism and properties
that optical interferometry and high-contrast imaging have in common: we first
focused on the notion of coherence, most often invoked in the sole context of
interferometry, since this technique directly aims at measuring it. The
fundamental coherence properties of astronomical sources however also make it
possible to explain how images acquired by telescopes form. They also explain
what information can be extracted from images dominated by diffraction
features. The link between the two techniques runs strong: the Van Cittert
Zernike theorem, used at the very heart of interferometry to relate the
measurements of complex visibilities to the properties of astrophysical sources
(refer to the lecture by Prof. Jean Surdej in this book) can be understood as a
Fourier-centric equivalent of the image - object convolution relation.

Equipped with this formal background, we took a closer look at high-contrast
imaging and the principles behind the optical techniques of pupil apodization
and coronagraphy that attempt to beat down the photon noise of the bright star
and improve the detectability of high-contrast sources in their vicinity.  We
now know that these solutions can only reduce the photon noise associated to
the static diffraction figure of the instrumental chain. In the presence of
residual aberrations, their performance rapidly degrades and their benefit
becomes marginal. State of the art extreme adaptive optics systems, manage to
bring the wavefront residual errors down to a few tens of nanometers, but
systematic biases, mostly associated to non-common path errors do survive over
long time-scales and limit the discovery potential of high-contrast imaging
instruments. Sophisticated post-processing techniques do manage to calibrate
some of these systematics and have considerably contributed to the direct
imaging of a few planetary systems featuring bright planets. There are still
orders of magnitude to overcome to directly image the large number of mature
planets theoretically within the grasp of ground based telescopes. Before
having to resort to post-processing, closed-loop feedback from the focal plane
while carrying out the observations seems like a reasonable way to compensate
for biases, and design systems that better answer the question: speckle or
planet? Going full circle back to the notion of coherence, the example of
iterative speckle nulling was described in higher details. Other techniques and
algorithms are also possible and may prove more efficient to implement as they
mature and adapt to the tough telescope environment. The robustness of the
speckle nulling approach however makes it an attractive next step in the
elimination of biases for high-contrast imaging.

%\bibliography{../../ms}

\end{document}